%% file: plar_draft.tex
\documentclass[10pt,journal,a4paper,compsoc]{IEEEtran}
\usepackage{amssymb,amsmath}
\usepackage{mathrsfs}
\usepackage{graphicx,amsfonts,listings,xcolor}
\usepackage[linesnumbered,ruled,vlined]{algorithm2e}
\let\oldnl\nl
\newcommand{\nonl}{\renewcommand{\nl}{\let\nl\oldnl}}

\usepackage{url}
\usepackage{threeparttable}
\usepackage{multirow}
\usepackage{subfigure}

\newtheorem {definition} {\bf Definition}[section]

\newtheorem {cor} {\bf Corollary}[section]

\newtheorem{Proof}{Proof}
\newtheorem{example}{Example}
\newcommand{\vect}[1]{\overset{\rightharpoonup}{#1}}
\newcommand\defeq{\mathrel{\overset{\makebox[0pt]{\mbox{\normalfont\tiny\sffamily def}}}{=}}}

\usepackage{color}
\definecolor{deepblue}{rgb}{0,0,0.5}
\definecolor{deepred}{rgb}{0.6,0,0}
\definecolor{deepgreen}{rgb}{0,0.5,0}

\newcommand{\Spark}{\textsc{Spark}}

\newcommand{\textFile}[1]{{\color{deepred}\textbf{textFile}({\color{deepblue}#1})}}
\newcommand{\map}[1]{{\color{deepred}\textbf{map}({\color{deepblue}#1})}}
\newcommand{\cache}{{\color{deepred}\textbf{cache}()}}

\newcommand{\reduceByKey}[1]{{\color{deepred}\textbf{reduceByKey}({\color{deepblue}#1})}}
\newcommand{\Sum}{{\color{deepred}\textbf{sum}()}}

\DeclareMathOperator*{\argmax}{arg\,max}
\DeclareMathOperator*{\argmin}{arg\,min}
\ifCLASSINFOpdf
\else
\fi

\begin{document}
%

\title{Parallel Large-Scale Attribute Reduction on Cloud Systems \thanks{This is an extended version of the paper presented at 2013 International Conference on Parallel and Distributed Computing, Applications and Technologies, 2013. An early version (in Chinese) can be found in Chapter 3 of the book \cite{Li2016}.}}

%

\author{Junbo~Zhang,
        Tianrui~Li,
        Yi~Pan
\thanks{Junbo~Zhang and Tianrui Li are with the School of Information Science and Technology, Southwest Jiaotong University, Chengdu 610031, China
(e-mail: jbzhang@my.swjtu.edu.cn, trli@swjtu.edu.cn);}
\thanks{Yi Pan is with the Department of Computer Science, Georgia State University, Atlanta, GA 30303, USA (e-mail: pan@cs.gsu.edu).}}

%
%

\markboth{Technical Report}
{Zhang \MakeLowercase{\textit{et al.}}: Parallel Large-Scale Attribute Reduction on Cloud Systems}
%


\maketitle

\begin{abstract}
The rapid growth of emerging information technologies and application patterns in modern society, \textit{e.g.}, Internet, Internet of Things, Cloud Computing and Tri-network Convergence, has caused the advent of the era of big data. Big data contains huge values, however, mining knowledge from big data is a tremendously challenging task because of data uncertainty and inconsistency. Attribute reduction (also known as feature selection) can not only be used as an effective preprocessing step, but also exploits the data redundancy to reduce the uncertainty. 
However, existing solutions are designed 1) either for a single machine that means the entire data must fit in the main memory and the parallelism is limited; 2) or for the Hadoop platform which means that the data have to be loaded into the distributed memory frequently and therefore become  inefficient. 
In this paper, we overcome these shortcomings for maximum efficiency possible, and propose a unified framework for \underline{P}arallel \underline{L}arge-scale \underline{A}ttribute \underline{R}eduction, termed PLAR, for big data analysis. 
PLAR consists of three components: 1) \underline{Gr}anular \underline{C}omputing (GrC)-based \textit{initialization}: it converts a decision table (\textit{i.e.} original data representation) into a granularity representation which reduces the amount of space and hence can be easily cached in the distributed memory:
2) \textit{model-parallelism}: it simultaneously evaluates all feature candidates and makes attribute reduction highly parallelizable; 
3) \textit{data-parallelism}: it computes the significance of an attribute in parallel using a MapReduce-style manner. 
We implement PLAR with four representative heuristic feature selection algorithms on \textsc{Spark}, and evaluate them on various huge datasets, including UCI and astronomical datasets, finding our method's advantages beyond existing solutions. 
\end{abstract}


%
\IEEEpeerreviewmaketitle

\section{Introduction}
Enormous amounts of data are generated every day with the amazing spread of computers and sensors in a wide-range of domains, including social media, search engines, insurance companies, health care organizations, financial industry and many others \cite{Li2015}.
Now we are in the era of big data, which is characterized by 5Vs \cite{Jin2015BDR}: 
1)~\textit{Volume} means the amount of data that needs to be managed is very huge; 
2)~\textit{Velocity} means that the speed of data update is very high;
3)~\textit{Variety} means that the data is varied in nature and there are many different types of data that need to be properly combined to make the most of the analysis; 
4)~\textit{Value} means high yield will be achieved if the big data is handled correctly and accurately; 
5)~\textit{Veracity} means the inherent uncertainty and ambiguity of data. 
Big data is currently a fast growing field both from an application and a research point of view. 
Since big data contains huge values, mastering big data means mastering resources. 
However, mining knowledge from big data is a tremendously challenging task because of data uncertainty and inconsistency. 
Attribute reduction, also known as feature selection in pattern recognition \& machine learning, can not only be used as an effective preprocessing step which reduces the complexity of handling big data, but also exploits the data redundancy to reduce the uncertainty from big data. 
It also helps people better understand the data by telling them which are key features, and has been attracted much attention in recent years \cite{Dash2003AI,Hu2010IToSMaCPBC,Qian2010AI,Hu2007PR,Qian2011PR,Yang2010}.  
Rough set theory, introduced by Pawlak in 1982, is a soft computing tool for dealing with inconsistent information in decision situations \cite{Pawlak1991,Pawlak2007IS,Pawlak2007ISb}, and plays an important role in the fields of
pattern recognition, feature selection and knowledge discovery \cite{Ziarko1999CotA}.
Attribute reduction in rough set theory provides a theoretic framework for consistency-based feature selection, which can retain the discernible ability of original features for the objects from the universe \cite{Qian2010AI}.

Attribute reduction from large data is an expensive preprocessing step. To accelerate this process, 
incremental techniques combined with traditional rough set based methods are widely researched \cite{Li2007KS,Zhang2012IJoAR,Cheng2011DKE,Qian2011PR,Qian2010AI,Hu2008IS,Liang2014IToKaDE}.
For example, Li et al. proposed an incremental method for dynamic attribute generalization, which can accelerate a heuristic process of attribute reduction by updating approximations incrementally.
Qian et al. introduced positive approximation, which is a theoretic framework for accelerating a heuristic process \cite{Qian2010AI}.
Zhang et al. presented a matrix-based incremental method for fast computing rough approximations.
Liang et al. developed a group of incremental feature selection algorithms based on rough sets \cite{Liang2014IToKaDE}. 

As these methods are still sequential and can not process big data with a cluster. 
MapReduce, by Google, is a popular parallel programming model and a framework for processing big data on certain kinds of distributable problems using a large number of computers, collectively referred to as a cluster \cite{Dean2008CotA}.
It can help arrange the application in the cluster easily.
Some MapReduce runtime systems were implemented, such as Hadoop \cite{White2012}, Twister \cite{Ekanayake2010}, Phoenix \cite{Talbot2011} and Mars \cite{He2008}, which all can help developers to parallelize traditional algorithms by using MapReduce model.
For example, Apache Mahout \cite{Owen2012} is machine learning libraries, and produces implementations of parallel scalable machine learning algorithms on Hadoop platform by using MapReduce. 

In the previous work, Zhang et al. developed a parallel algorithm for computing rough set approximations based on MapReduce \cite{Zhang2012IS}.
Based on that, Zhang et al. proposed a parallel rough set based knowledge acquisition method using MapReduce \cite{Zhang2012}. 
Afterwards, Qian et al. presented a parallel attribute reduction algorithm based on MapReduce \cite{Qian2014IS}. 
However, all of these existing parallel methods make use of the classical MapReduce framework and are implemented on the Hadoop platform \cite{White2012}. 

In this paper, we develop a computational method for attribute reduction that meets all the following criteria: 
1) it can support devise attribute significance measures; 
2) it is highly parallelizable; 
and 3) it can evaluate feature candidates simultaneously without evaluating one by one. 
Our main contributions are four-folds: 
\begin{itemize}
\item We propose a novel parallel method, called PLAR, for large-scale attribute reduction, which supports data parallelism and model parallelism. 
\item Our method is a unified framework for attribute reduction, which means that various attribute reduction algorithms can be integrated into ours. 
\item We implement our method, PLAR, on a general-purpose in-memory dataflow system, \Spark{} \cite{Spark}, and conduct extensive experimental evaluation. In contrast, the most existing methods \cite{Qian2010AI,Qian2014IS} are designed 1) either for a single machine which means that the entire data must fit in the main memory and the parallelism is limited; 2) or for Hadoop which means that the data have to be loaded into the distributed memory frequently. 
\item We test on the large sample size \& high-demensional data, SDSS, with 320,000 samples and 5201 features, which is a real-world astronomical dataset. It takes about 7,000 seconds for one iteration. 
\end{itemize}


The rest of this paper is organized as follows.
Section~\ref{sec:P} gives the elementary background introduction to feature selection and \Spark{}. 
In Section~\ref{sec:Pa}, we propose a parallel framework for feature selection and a unified representation of attributes' significance evaluation functions. 
We describe our implementation in Section~\ref{sec:Im}, and present the experimental analysis in Section~\ref{sec:E}. 
The paper ends with conclusions and future work in Section~\ref{sec:C}. 

\section{Preliminaries}\label{sec:P}
In this section, we first review a unified framework for original feature selection as well as four representative significance measures of attributes, then introduce the distributed data processing platform used in this paper.

\subsection{A Unified Framework for Feature Selection} \label{sec:sub:EF}
\begin{figure}[!htbp]
\begin{center}
\includegraphics*[width=0.45\textwidth]{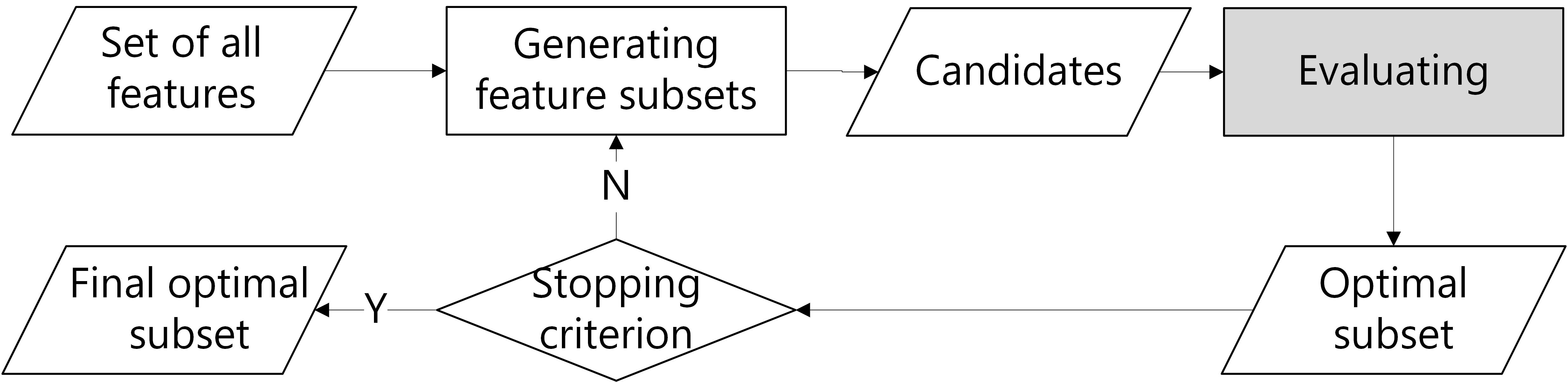}
\end{center}
\caption{A unified framework for feature selection} \label{fig:AR}
\end{figure}

Figure~\ref{fig:AR} shows a unified framework for sequential feature selection.
The first step is generating several attribute subsets. The second step is to employ an attribute significance measure to evaluate all generated candidates, and outputs the current optimal feature subset. The third step is checking whether the stopping criterion (\textit{\textit{e.g.}}, number of attributes) is satisfied, a) if yes, output the current optimal feature subset as final optimal subset; b) otherwise, it goes back the first step, generating candidates then evaluating until the selected feature subset meets the stopping criterion.

In this framework, the key component is the evaluating and almost the computational work comes from this step. Hence, the efficiency of feature selection depends on evaluating. The main contribution of this paper is to extend this framework and propose a parallel framework to accelerate the whole process of evaluating.
For the evaluation of attributes, there are mainly two general methods: a) \textit{wrapper} which employs a learning algorithm (\textit{\textit{e.g.}} support vector machine) to evaluate; b) \textit{filter} which measures the attributes' significance with a metric, including
information gain \cite{Liang2002IJoUFaKS}, distance \cite{Kira1992}, dependency \cite{Modrzejewski1993} and consistency \cite{Dash2003AI,Qian2008Fsas}.
The representative significance measures of attributes used in this paper are all based on rough set theory. It would also be interesting to study how to use our proposed framework to scale up on other attribute reduction algorithms.

\subsubsection{Heuristic attribute reduction algorithm}
To select optimal feature subset efficiently and effectively, many heuristic attribute reduction algorithms were proposed during pasting two decades  \cite{Liang2002IJoGS,Liang2002IJoUFaKS,Qian2008IJoUFaKS,Slezak2002FIa}, most of which make use of forward search strategy.
In each forward heuristic attribute reduction algorithm, starting with the attributes (called \textit{core}) with the satisfied inner importance (\textit{\textit{e.g.}}, greater than a threshold), it takes the attribute with the maximal outer significance into the feature subset iteratively until the selected feature subset meets the stopping criterion, and finally we can get an attribute reduct \cite{Qian2010AI}. 

Let $S=(U, C\cup D)$ be a decision table, $B \subseteq C$. We denote the \textit{inner} and \textit{outer} importance measures of an attribute $a$ as $Sig^{inner}_{\Delta}(a, B, D)$ and $Sig^{outer}_{\Delta}(a, B, D)$, respectively. In general, \textit{inner} is used to measure an internal attribute $a \in B$ by removing it from $B$; \textit{outer} is employed to measure an external attribute $a$ by adding it into $B$. Four representative metrics $\Delta \in \{PR, SCE, LCE, CCE\}$ for measuring the attribute's importance will be introduced in Section~\ref{sec:significance}. 
Based on these notation, we give the unified definition for computing the attribute core and selecting the best attribute from the candidates as follows. 

\begin{definition}[Attribute Core] 
Let $S = (U,C \cup D)$ be a decision table. $Sig^{inner}_{\Delta}(a,C,D)$ is an inner importance measure of an attribute $a$, and $\epsilon$ is a threshold. The core of attributes, termed $Core$, is defined as
\begin{equation}
Core=\{a| Sig^{inner}_{\Delta}(a,C,D)>\epsilon, a\in C\}
\end{equation}
\end{definition}

\begin{definition}[Optimal Attribute]\label{defi:a_opt} 
Let $S = (U,C \cup D)$ be a decision table, $B\subseteq C$. $Sig^{outer}_{\Delta}(a,B,D)$ is an outer importance measure of an attribute $a$.
The optimal attribute $a_{opt}$ is defined as
\begin{equation}
a_{opt}=\argmax\limits_{a\in C\setminus B}\{Sig^{outer}_{\Delta}(a,B,D)\}
\end{equation}
\end{definition}

\begin{figure}[!htbp]
\begin{center}
\includegraphics*[width=0.35\textwidth]{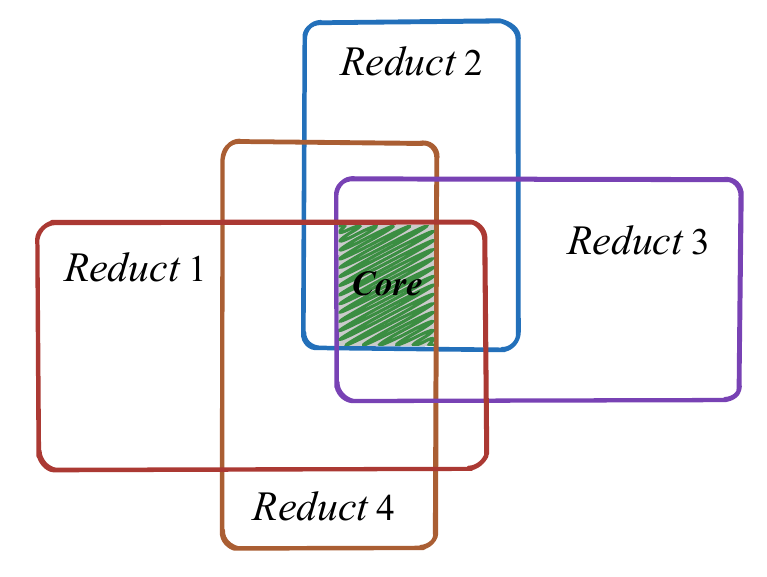}
\end{center}
\caption{The relationship between \textit{core} and \textit{reduct}} \label{fig:core_and_reduct}
\end{figure}

Figure~\ref{fig:core_and_reduct} shows the relationship between \textit{core} and \textit{reduct}. In detail, a general forward heuristic attribute reduction algorithm can be written as follows. 

\begin{algorithm}
\caption{A general forward heuristic attribute reduction algorithm}\label{alg:seqAR}
\KwIn{A decision table $S=(U, C\cup D)$, attribute importance measure metric $\Delta$, threshold $\epsilon$}
\KwOut{An attribute reduct $\mathcal R$}

  $Core=\{a| Sig^{inner}_{\Delta}(a,C,D)>\epsilon, a\in C\}$ \tcp*{$Sig^{inner}_{\Delta}(a,C,D)$ is the inner importance measure of the attribute $a$}
  $\mathcal R \longleftarrow Core$  \;
  \While{stopping criterion not met $\&$ $C\setminus\mathcal R \ne \emptyset$}{
  $a_{opt}=\argmax\limits_{a\in C\setminus\mathcal R}\{Sig^{outer}_{\Delta}(a,\mathcal R,D)\}$ \tcp*{$Sig^{outer}_{\Delta}(a,\mathcal R,D)$ is the outer importance measure of the attribute $a$}
  $\mathcal R \longleftarrow \mathcal R \cup \{a_{opt}\}$\;
  }
  \Return $\mathcal R$
\end{algorithm}

\subsubsection{Representative significance measures of attributes}\label{sec:significance}
\noindent For efficient attribute reduction, many heuristic attribute reduction methods have been developed in rough set theory  \cite{Liang2002IJoGS,Liang2002IJoUFaKS,Qian2008IJoUFaKS,Slezak2002FIa}.
Further, from the viewpoint of heuristic functions, Qian et al. classified these attribute reduction methods into four categories: positive-region reduction, Shannon's conditional entropy reduction, Liang's conditional entropy reduction and combination conditional entropy reduction \cite{Qian2010AI}. We here also only focus on these four representative attribute reduction methods.

Given a decision table $S = (U, C \cup D)$, $B \subseteq C$, the condition partition $U/B = \{E_1, E_2, \cdots, E_e\}$ and the decision partition $U/D=\{D_1,D_2,\cdots,D_m\}$ can be obtained. Through these notations, we briefly review four types of significance measures of attributes as follows.

\noindent(1) Positive-region based method
\begin{definition}[PR]\label{def:PR}
Let $S=(U, C \cup D)$ be a decision table, $B \subseteq C$. The dependency degree of $D$ to $B$ is defined as

$\gamma_B(D) = \frac{|POS_B(D)|}{|U|}$,\\
where
$POS_B(D) = \bigcup\limits_{i=1}^m \{E_i \in U/B : X_i \subseteq Y_1 \vee X_i \subseteq Y_2 \vee \cdots \vee X_i \subseteq Y_n \} = \bigcup\limits_{i=1}^m \{X_i \in U/B : |X_i/D| = 1 \} $.
\end{definition}

\begin{definition}[PR significance]
Let $S=(U, C \cup D)$ be a decision table, $B \subseteq C$.
The inner and outer significance measure of $a$ based on PR, denoted by $Sig_{PR}^{inner}(a, B, D)$ and $Sig_{PR}^{outer}(a, B, D)$, are respectively defined as
\small
\begin{eqnarray}
 Sig_{PR}^{inner}(a, B, D) =\gamma_B(D) - \gamma_{B\backslash\{a\}}(D), \forall a \in B \nonumber \\
 Sig_{PR}^{outer}(a, B, D) =\gamma_{B \cup \{a\}}(D) - \gamma_B(D), \forall a \in C\backslash B \nonumber
\end{eqnarray}
\end{definition}
\noindent(2)~Shannon's conditional entropy based method (SCE)

\begin{definition}[SCE]
Shannon's conditional entropy of $D$ with respect to $B$ is defined as
\begin{equation}
\mathcal H(D|B) = - \sum\limits_{i=1}^e p(E_i) \sum\limits_{j=1}^m p(D_j|E_i)\log(p(D_j|E_i))
\end{equation}
where $p(E_i) = \frac{|E_i|}{|U|}$, $p(D_j|E_i) = \frac{|E_i \cap D_j|}{|E_i|}$.
\end{definition}

\begin{definition}[SCE significance]
Let $S=(U, C \cup D)$ be a decision table, $B \subseteq C$.
The inner and outer significance measure of $a$ based on SCE, denoted by $Sig_{SCE}^{inner}(a, B, D)$ and $Sig_{SCE}^{outer}(a, B, D)$, are respectively defined as
\small
\begin{eqnarray}
 Sig_{SCE}^{inner}(a, B, D) =\mathcal H(D|B \backslash \{a\}) - \mathcal H(D|B), \forall a \in B \nonumber \\
 Sig_{SCE}^{outer}(a, B, D) =  \mathcal H(D|B) - \mathcal H(D|B \cup \{a\}), \forall a \in C\backslash B \nonumber
\end{eqnarray}
\end{definition}

\noindent(3)~Liang's conditional entropy based method (LCE)

\begin{definition}[LCE]
Liang's conditional entropy of $D$ with respect to $B$ is defined as
\begin{equation}
\mathcal H_L(D|B) = \sum\limits_{i=1}^e \sum\limits_{j=1}^m \frac{|D_j \cap E_i|}{|U|} \frac{|D_j^c \setminus E_i^c|}{|U|}
\end{equation}
where $E^c$ means the complement of the set $E$.
\end{definition}

\begin{definition}[LCE significance]
Let $S=(U, C \cup D)$ be a decision table, $B \subseteq C$.
The inner and outer significance measure of $a$ based on LCE, denoted by $Sig_{LCE}^{inner}(a, B, D)$ and $Sig_{LCE}^{outer}(a, B, D)$, are respectively defined as
\small
\begin{eqnarray}
 Sig_{LCE}^{inner}(a, B, D) = \mathcal H_L (D|B \backslash \{a\}) - \mathcal H_L(D|B), \forall a \in B \nonumber \\
 Sig_{LCE}^{outer}(a, B, D) =  \mathcal H_L (D|B) - \mathcal H_L(D|B \cup \{a\}), \forall a \in C\backslash B \nonumber
\end{eqnarray}
\end{definition}

\noindent(4)~Combination conditional entropy based method (CCE)
\begin{definition}[CCE]
Combination conditional entropy of $D$ with respect to $B$ is defined as

\begin{equation}
\mathcal H_Q(D|B)=\sum\limits_{i=1}^e \left( \frac{|E_i|}{|U|}\frac{C_{|E_i|}^2}{C_{|U|}^2} -
    \sum\limits_{j=1}^m \frac{|E_i \cap D_j|}{|U|}\frac{C_{|E_i \cap D_j|}^2}{C_{|U|}^2} \right)
\end{equation}
where $C_{|E_i|}^2= \frac{|E_i|\times(|E_i|-1)}{2}$ denotes the number of the pairs of the objects which are not distinguishable each other in the equivalence class $E_i$.
\end{definition}
\begin{definition}[CCE significance]
Let $S=(U, C \cup D)$ be a decision table, $B \subseteq C$.
The inner and outer significance measure of $a$ based on CCE, denoted by $Sig_{CCE}^{inner}(a, B, D)$ and $Sig_{CCE}^{outer}(a, B, D)$, are respectively defined as
\small
\begin{eqnarray}
 Sig_{CCE}^{inner}(a, B, D) = \mathcal H_Q(D|B \backslash \{a\}) - \mathcal H_Q(D|B), \forall a \in B \nonumber \\
 Sig_{CCE}^{outer}(a, B, D) =  \mathcal H_Q(D|B) - \mathcal H_Q(D|B \cup \{a\}), \forall a \in C\backslash B \nonumber
\end{eqnarray}
\end{definition}

Intuitively, these four significance measures of attributes are listed in Table~\ref{tab:4Methods}.
To keep the notation consistent, we define $\gamma(D|B) \defeq - \gamma_B(D)$,
hence, $\gamma(\cdot)$, $\mathcal H(\cdot)$, $\mathcal H_L(\cdot)$ and $\mathcal H_Q(\cdot)$ can be written as a unified form, \textit{\textit{i.e.}}, $\Theta(\cdot)$. And, the inner and outer importance measures can be computed by $Sig_{\Delta}^{inner} = \Theta(D|B \backslash \{a\}) -  \Theta(D|B)$ and $Sig_{\Delta}^{outer} = \Theta(D|B) - \Theta(D|B \cup \{a\})$
, respectively.
Thus, it is easy to see that the attribute's significance measure can be transformed to the computation of the $\Theta(D|B)$.
\begin{table*}[htbp]
\caption{Four representative significance measures of attributes}\vspace*{-15pt}
\label{tab:4Methods}
\begin{center}
\begin{tabular}{clcc}
\hline
      $\Delta$  & $\Theta(D|B)$ & $Sig^{inter}_\Delta(a,B,D)$ & $Sig^{outer}_\Delta(a,B,D)$\\
      (Metric)&  & ($\forall a \in B$) & ($\forall a \in C \setminus B$) \\
\hline
    PR  & $\gamma(D|B) \defeq - \gamma_B(D) = - \frac{|POS_B(D)|}{|U|}$  & $\gamma(D|B\backslash \{a\}) - \gamma(D|B)$& $\gamma(D|B) - \gamma(D|B\cup \{a\})$ \\
   SCE   &  $\mathcal H(D|B) = - \sum\limits_{i=1}^e p(E_i) \sum\limits_{j=1}^m p(D_j|E_i)\log(p(D_j|E_i))$
   & $\mathcal H(D|B\setminus \{a\}) - \mathcal H(D|B)$ & $\mathcal H(D|B) - \mathcal H(D|B\cup \{a\})$ \\
  LCE    & $\mathcal H_L(D|B) = \sum\limits_{i=1}^e \sum\limits_{j=1}^m \frac{|D_j \cap E_i|}{|U|} \frac{|D_j^c - E_i^c|}{|U|}$
  &$\mathcal H_L(D|B\setminus \{a\}) - \mathcal H_L(D|B)$ & $\mathcal H_L(D|B) - \mathcal H_L(D|B\cup \{a\})$\\
  CCE    &  $\mathcal H_Q(D|B)=\sum\limits_{i=1}^e \left( \frac{|E_i|}{|U|}\frac{C_{|E_i|}^2}{C_{|U|}^2} -
    \sum\limits_{j=1}^m \frac{|E_i \cap D_j|}{|U|}\frac{C_{|E_i \cap D_j|}^2}{C_{|U|}^2}\right)$
    &$\mathcal H_Q(D|B\setminus \{a\}) - \mathcal H_Q(D|B)$ & $\mathcal H_Q(D|B) - \mathcal H_Q(D|B\cup \{a\})$ \\
\hline
\end{tabular}
\end{center}
\end{table*}

\subsection{\textsc{Spark}}
We propose a unified framework for parallel large-scale attribute reduction, which is distributed in nature. Therefore, in principle, it can be implemented on any distributed data processing platforms (\textit{\textit{e.g.}} \textsc{Hadoop}, \textsc{Spark}). In our implementation, \textsc{Spark} \cite{Spark} is chosen because of its properties: a) \textsc{Spark} has currently  developed into a full-fledged, general-purpose distributed computation engine for large-scale data processing, and facilitates in-memory cluster computing, which is essential for iterative algorithms; b) it provides easy-to-use operations to build parallel, distributed and fault-tolerant applications; c) it is implemented in Scala, programming APIs in Scala, Java, R, and Python makes \textsc{Spark} much more accessible to a range of data scientists who can take fast and full advantage of the Spark engine.

At a high level, every \Spark{} application consists of a driver program that runs the user’s main function and executes various parallel operations on a cluster. The main abstraction \Spark{} provides is a Resilient Distributed Dataset (RDD) \cite{Zaharia2012} that allows applications to keep data in the shared memory of multiple machines and can be operated on in parallel. RDDs are fault-tolerant since \Spark{} can automatically recover lost data.
Formally, an RDD is a read-only, partitioned collection of records, where
two types of operations over RDDs are available: \textit{transformations} which create a new RDD from an existing one, and \textit{actions} which return a value to the driver program after running a computation on the RDD.
\textit{Transformations} used in our implementations mainly include \texttt{map}, \texttt{flatMap}, and \texttt{reduceByKey}. Specifically, \texttt{map} passes each dataset element through a function and returns a new RDD representing the results which is typically one-to-one mapping of the input RDD, and \texttt{flatMap} constructs a one-to-many of the input; \texttt{reduceByKey} works on an RDD of key-value pairs and generates a new RDD of key-value pairs where the values for each key are aggregated using the given reduce function. These operations are similar to the \textit{map} and \textit{reduce} operations in the traditional \textsc{MapReduce} \cite{Dean2008CotA} framework.
We use two actions: \texttt{collect} and \texttt{sum}, the former returns all elements of an RDD and the latter returns the sum of all elements of an RDD.
In additional to these operations, one can call \texttt{cache} or \texttt{persist} to indicate which RDD to be reused in future and \Spark{} will cache persistent RDDs in memory, which radically reduces the computational time.

\section{Parallel Large-Scale Attribute Reduction}\label{sec:Pa}
In this section, we first propose a hybrid parallel framework for attribute reduction, then present a unified representation of evaluation functions, finally give a \Spark{}-based algorithm for large-scale attribute reduction.

\subsection{A Parallel Framework for Attribute Reduction}
We here give a parallel framework for attribute reduction. 
It mainly has the following stages: 1)~Generating candidates using a certain search strategy (\textit{\textit{e.g.}}, heuristic search in this paper). In general, this stage generates a pool of attributes. We propose to leverage a multi-processing method to parallelize the processing of evaluation of multiple attributes. As this procedure is model-related, we call the method as Model-Parallelism (MP). 
2)~With lunching a pool of worker processes, each of which is used to compute the significance of an attribute in \Spark{}. Here, $\mathcal R$ is the attribute reduct in the current loop. This processing makes use of \Spark{} to parallelize the processing of the computation of an attribute's importance. Hence it is named as Data-Parallelism (DP). 
3)~After computing all features'  significance, the optimal feature is selected according to the best significance and inserted into the $\mathcal R$. 
Then, it goes into next loop. 
By combining MP and DP, our proposed framework formally supports both model and data parallelism. Therefore, we term this framework model-and-data-parallelism (MDP). 

\begin{figure}[!htbp]
\begin{center}
\includegraphics*[width=0.5\textwidth]{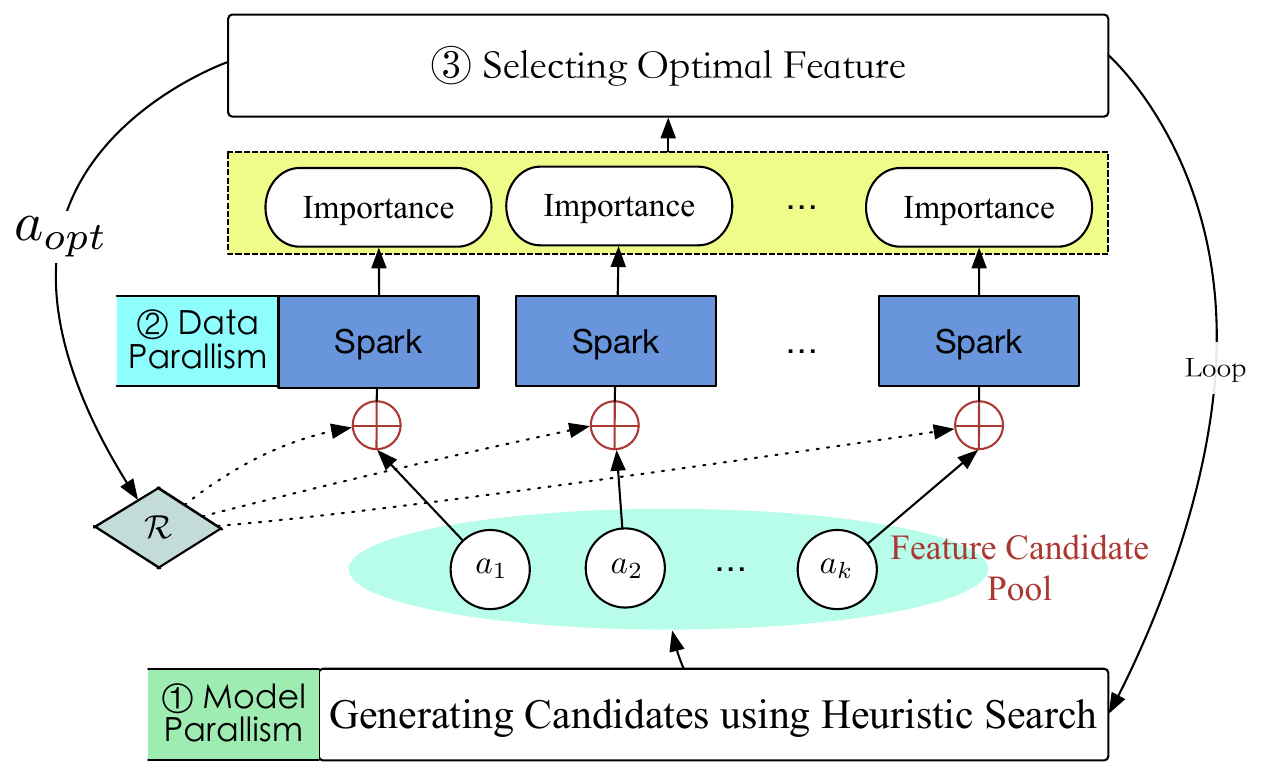}
\end{center}
\caption{A Parallel Framework for Attribute Reduction
} \label{fig:para_framework}
\end{figure}

\subsection{A Unified Representation of Evaluation Functions}\label{sec:EF}
In this section, we first introduce the simplification and decomposition of evaluation functions which is a unified view for computing evaluation functions' value. Based on it, we give a MapReduce-based method. 

\subsubsection{Simplification and Decomposition of Evaluation Functions}

\begin{cor}\label{cor:pr:divide}
Let $S = (U,C \cup D)$ be a decision table, $B \subseteq C$. $U/B = \{E_1, E_2, \cdots, E_e\}$ and $U/D=\{D_1,D_2,\cdots,D_m\}$ are  condition and decision partitions, respectively.

The positive region $POS_B(D)$ is calculated as follows
\begin{eqnarray}
POS_B(D) 
         = \bigcup\limits_{i=1}^e \{E_i \in U/B : |E_i/D| = 1 \}
\end{eqnarray}
\end{cor}
where $E_i/D$ is the decision partition of $E_i$.
\begin{Proof}
Given $U/D=\{D_1, D_2, \cdots, D_m\}$, $\forall E_i \in U/B$. 
\emph{i)}~If $\exists D_j \in U/D$, $E_i \subseteq D_j$, then $E_i \subseteq \underline R(D_j) \subseteq POS_B(D)$, and $|E_i/D|=1$; 
\emph{ii)}~If $\nexists D_j \in U/D$, $E_i \subseteq D_j$, then $E_i \not\subseteq POS_B(D)$ and $|E_i/D|>1$.

To sum up, we have
\begin{flalign*}
&&POS_B(D)&=\bigcup\limits_{j=1}^m \left( \bigcup\limits_{i=1}^e \{E_i \in U/B: E_i \subseteq D_j\} \right) \nonumber \\
&& &= \bigcup\limits_{i=1}^e \left( \bigcup\limits_{j=1}^m \{E_i \in U/B: E_i \subseteq D_j\} \right) \nonumber \\
&& &= \bigcup\limits_{i=1}^e \{E_i \in U/B : E_i \subseteq D_1 \vee E_i \subseteq D_2 \vee \nonumber \\ &&& \quad\  \cdots \vee E_i \subseteq D_m \} \nonumber \\
&& &= \bigcup\limits_{i=1}^e \{E_i \in U/B : |E_i/D| = 1 \}.  \qquad \Box
\end{flalign*}
\end{Proof}

\begin{definition}[Multiset]\label{defi:cond_deci_part}
Let $S = (U,C \cup D)$ be a decision table, $B \subseteq C$. $U/B = \{E_1, E_2, \cdots, E_e\}$ and $U/D=\{D_1,D_2,\cdots,D_m\}$ are  condition and decision partitions, respectively. 
$\forall E_i\in U/B$, the multiset of the decision partition of $E_i$,  denoted by $E_i//D$, is defined as 
\begin{eqnarray}
E_i//D=\{D_{i1}, D_{i2}, \cdots,D_{im}\}
\end{eqnarray}
where $D_{ij}=E_i \cap D_j$, $\forall D_j\in U/D$. 
\end{definition}

\begin{cor}\label{cor:cond_deci_part}
Let $S = (U,C \cup D)$ be a decision table, $B \subseteq C$. $U/B = \{E_1, E_2, \cdots, E_e\}$ and $U/D=\{D_1,D_2,\cdots,D_m\}$ are condition and decision partitions of $U$, respectively. 
$E_i/D=\{Y_{i1}, Y_{i2}, \cdots, Y_{il}\}$ is the decision partition of $E_i$, $\forall E_i\in U/B$. 
Then, the multiset of the decision partition of $E_i$, $E_i//D = \{D_{i1}, D_{i2}, \cdots,D_{im}\}$ can be computed by 
\begin{eqnarray}
D_{ij} = \left\{ {\begin{array}{*{20}c}
   Y, & \exists ! Y \in E_i/D, \vect Y = \vect D_{ij}\\
   \emptyset, & else \\
\end{array}} \right.
, \forall j = 1,\cdots, m \nonumber
\end{eqnarray}
where $\vect Y$ and $\vect D_{ij}$ mean the labels of the decision class $Y$ and $D_{ij}$. 
\end{cor}
\begin{Proof}
If $\exists! Y \in E_i/D, \vect Y = \vect D_{ij}$, 
because $\vect D_{ij} = \vect D_{j}$, 
therefore $\vect Y = \vect D_{j}$. 
As $Y\subseteq E_i \subseteq U$ and $D_j\in U/D$, according to the definition of equivalence partition \cite{Pawlak2007ISa}, we have $Y \subseteq D_j$. 
Thus, $Y=E_i \cap D_j = D_{ij}$, that is $D_{ij}=Y$.  \hfill$\Box$
\end{Proof}

\noindent(1) Decomposition of PR-based Evaluation Function 

According to Corollary~\ref{cor:pr:divide}, we have 
\begin{eqnarray}
  \gamma(D|B) = - \frac{|POS_B(D)|}{|U|} 
   = \sum\limits _{i=1}^e \left( - \frac{|E_i|sgn_{PR}(E_i)}{|U|} \right) \nonumber
\end{eqnarray}
where  $sgn_{PR}(E_i) = \left\{ {\begin{array}{*{20}c}
   1, & |E_i/D| = 1 \\
   0, & else \\
\end{array}} \right.$.\\

\noindent(2) Decomposition of SCE-based Evaluation Function

According to Corollary~\ref{cor:cond_deci_part},  we have
\begin{eqnarray}
  \mathcal H(D|B) &=& - \sum\limits_{i=1}^e p(E_i) \sum\limits_{j=1}^m p(D_j|E_i)\log(p(D_j|E_i)) \nonumber \\
   &=& - \sum\limits_{i=1}^e \frac{|E_i|}{|U|} \sum\limits_{j=1}^m \frac{|E_i \cap D_j|}{|E_i|}\log \left(\frac{|E_i \cap D_j|}{|E_i|}\right) \nonumber \\
   &=&  \sum\limits_{i=1}^e \left ( - \frac{1}{|U|} \sum\limits_{j=1}^m |D_{ij}|\log \frac{|D_{ij}|}{|E_i|} \right )\nonumber
\end{eqnarray}

\noindent(3) Decomposition of LCE-based Evaluation Function

According to Corollary~\ref{cor:cond_deci_part}, we have
\begin{eqnarray}
  \mathcal H_L(D|B) &=& \sum\limits_{i=1}^e \sum\limits_{j=1}^m \frac{|D_j \cap E_i|}{|U|} \frac{|D_j^c - E_i^c|}{|U|} \nonumber \\
   &=& \sum\limits_{i=1}^e \sum\limits_{j=1}^m \frac{|D_j \cap E_i|}{|U|} \frac{|E_i - D_j|}{|U|}  \nonumber \\
   &=& \sum\limits_{i=1}^e  \left (  \sum\limits_{j=1}^m \frac{|D_j \cap E_i|}{|U|} \frac{|E_i| - |D_j \cap E_i|}{|U|}  \right )\nonumber \\
  &=& \sum\limits_{i=1}^e  \left (  \sum\limits_{j=1}^m \frac{|D_{ij}|(|E_i| - |D_{ij}|)}{|U|^2}  \right )\nonumber 
\end{eqnarray}


\noindent(4) Decomposition of CCE-based Evaluation Function

According to Corollary~\ref{cor:cond_deci_part}, we have 
\begin{eqnarray*}
  \mathcal H_Q(D|B) &=& \sum\limits_{i=1}^e \left ( \frac{|E_i|}{|U|}\frac{C_{|E_i|}^2}{C_{|U|}^2} -
    \sum\limits_{j=1}^m \frac{|E_i \cap D_j|}{|U|}\frac{C_{|E_i \cap D_j|}^2}{C_{|U|}^2} \right) \nonumber \\
   &=& \sum\limits_{i=1}^e \left ( \frac{|E_i|}{|U|}\frac{|E_i|\times(|E_i|-1)}{C_{|U|}^2} - \right. \nonumber \\
   && \left. \sum\limits_{j=1}^m \frac{|E_i \cap D_j|}{|U|}\frac{|E_i \cap D_j|\times(|E_i \cap D_j|-1)}{C_{|U|}^2} \right)  \nonumber \\
    &=& \sum\limits_{i=1}^e \left( \frac{|E_i|^2\times(|E_i|-1)}{|U|C_{|U|}^2} - \right. \nonumber \\ && \left.
     \sum\limits_{j=1}^m \frac{|E_i \cap D_j|^2\times(|E_i \cap D_j|-1)}{|U|C_{|U|}^2}\right)  \nonumber \\
&=& \sum\limits_{i=1}^e \left( \frac{|E_i|^2\times(|E_i|-1)}{|U|C_{|U|}^2} - \right. \nonumber \\
  &&  \left.  \sum\limits_{j=1}^m \frac{|D_{ij}|^2\times(|D_{ij}|-1)}{|U|C_{|U|}^2}\right)  \nonumber \\
\end{eqnarray*}

According to the above corollaries and transformation, these four evaluation functions can be written as a unified form as follows
\begin{equation}
  \Theta(D|B) = \sum\limits _{i=1}^e \theta(S_i)，
\end{equation}
where $S_i \defeq (E_i, D)$, four evaluation functions' sub function $\theta$ are shown in Table~\ref{tab:4Methods:Par}. 

\begin{table}[htbp]
\caption{Sub-Function of  Evaluation Function $\theta$}\vspace*{-15pt}
\label{tab:4Methods:Par}
\begin{center}
\begin{tabular}{cl}
\hline
$\Delta$ & $\theta(S_i)$ \\
\hline
    PR  & $-\frac{|E_i|sgn_{PR}(E_i)}{|U|}$ \\
   SCE  &  $- \frac{1}{|U|} \sum\limits_{j=1}^m |D_{ij}|\log \frac{|D_{ij}|}{|E_i|}$ \\
  LCE   & $\sum\limits_{j=1}^m \frac{|D_{ij}|(|E_i| - |D_{ij}|)}{|U|^2}$ \\
  CCE   & $\frac{|E_i|^2\times(|E_i|-1)}{|U|C_{|U|}^2} -
     \sum\limits_{j=1}^m \frac{|D_{ij}|^2\times(|D_{ij}|-1)}{|U|C_{|U|}^2}$ \\
\hline
\end{tabular}
\end{center}
\end{table}

\subsubsection{MapReduce-based Method}
\begin{figure}[!htbp]
\begin{center}
\includegraphics*[width=0.45\textwidth]{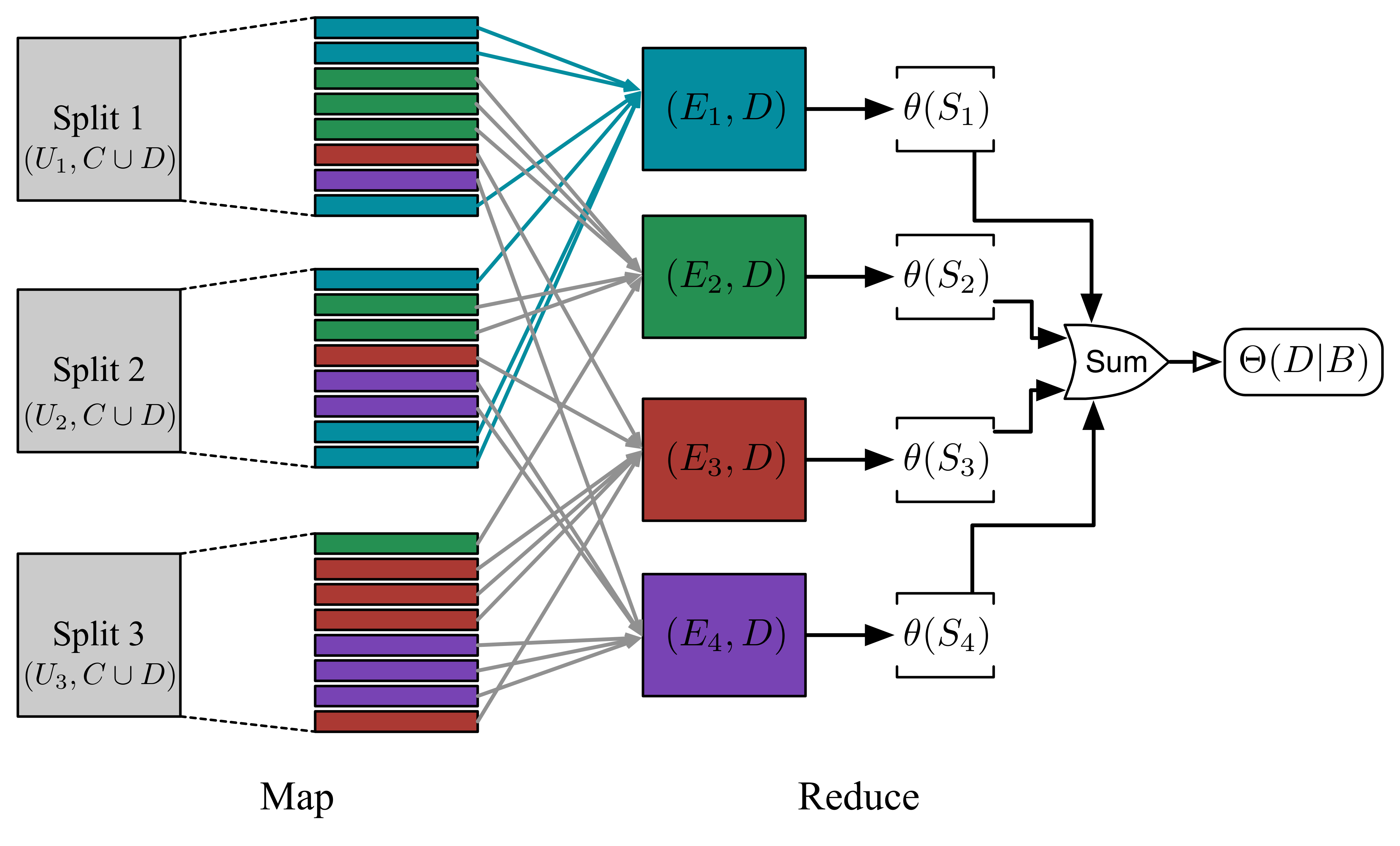}
\end{center}
\caption{Computing attribute significance using MapReduce} \label{fig:mr}
\end{figure}

According to the aforementioned simplification and decomposition, we describe the basic MapReduce-based method below, as shown in Figure~\ref{fig:mr}. The detailed \Spark{}-based algorithm will be introduced in Section~\ref{sec:Im}. Note that the basic parallel idea is similar, but \Spark{} provides much more exhaustive API beyond the native \textsc{Map} and \textsc{Reduce} functions. 

\begin{itemize}
\item \textsc{Map} phase: each Map worker reads data split $U_k$ from distributed file system (\textit{e.g.} HDFS), and then mapping its elements into key-value pair $(\vect x_B, \vect x_D)$. The main function of this processing is dividing $U_k$ into equivalence classes w.r.t. the attribute set $B$.
\item \textsc{Reduce} phase: each Reduce worker receives a group of data, whose key and value are $\vect {E_i}_B$ and $S_i=(E_i, D)$ respectively. Then, it computes the sub-evaluation function's value $\theta (S_i)$. 
\item \textsc{Sum} phase: finally, the main function collects the results from all Reduce workers, and calculates the sum over these collected values that is the evaluation function's value $\Theta (D|B)$. 
\end{itemize}

\subsection{Granularity Representation}\label{sec:GrC}
We propose to employ a granularity representation method to accelerate the processing procedure of feature selection. 
From the perspective of Granular Computing (GrC), $U$ in a decision table can be partitioned into different-scale (namely granularity) disjoint sets by different attribute sets. 
We first define the granularity representation of a decision table below. 

\begin{definition}[Granularity Representation]
Let $S = (U, A)$ be a decision table where $A = \{a_1, \cdots, a_{|A|}\}$. 
$U/A$ is a partition over $U$ w.r.t. $A$. The granularity representation of $S$ w.r.t. $A$, denoted as $G^{(A)}$, defined as follows, 
\begin{equation}
G^{(A)} = \{(\vect E_A, |E_A|) : E_A\in U/A\}
\end{equation}
where $\vect E_A \defeq \langle v_{a_1}, \cdots, v_{a_{|A|}}\rangle$ is the feature vector representation of the equivalence class $E_A$ w.r.t. $A$, and $v_{a_i}$ is the value on the attribute $a_i$, $\forall a_i \in A$ ; $|E_A|$ denotes the cardinality of $E_A$. 
\end{definition}

\begin{example}
A decision table $S=(U, C\cup D)$ is given in Table~\ref{tab:example}, where $C=\{a_1, a_2\}$ is the conditional attribute set and $D$ is the decision set. Assume that $A=C\cup D$, the partition over $U$ w.r.t. $A$ can be computed as $U/A=\{E_1, E_2, \cdots, E_5\}$, where $E_1 = \{x_1, x_2\}$, $E_2 = \{x_3\}$, $E_3 = \{x_4, x_5, x_6\}$, $E_4 = \{x_7\}$, and $E_5 = \{x_8\}$. Taking $E_1$ for example, its feature vector is $\vect{E_1} = \langle 0,0,Y\rangle$ and the associated cardinality is $|E_1| = 2$. The more details of the granularity representation can be found in Table~\ref{tab:gr}. 
\begin{table}[!htbp]
\caption{A decision table $S$}\vspace*{-10pt}
\label{tab:example}
\begin{center}
\begin{tabular}{clcc}
\hline
Object  & $a_1$ & $a_2$ & $D$ \\
\hline
$x_1$  & 0 & 0 & Y \\
$x_2$  & 0 & 0 & Y \\
$x_3$  & 0 & 0 & N \\
$x_4$  & 0 & 1 & Y \\
$x_5$  & 0 & 1 & Y \\
$x_6$  & 0 & 1 & Y \\
$x_7$  & 1 & 0 & N \\
$x_8$  & 1 & 1 & Y \\
\hline
\end{tabular}
\end{center}
\end{table}

\begin{table}[!htbp]
\caption{Granularity representation $G^{(A)}$, $A=\{a_1, a_2, D\}$}\vspace*{-10pt}
\label{tab:gr}
\begin{center}
\begin{tabular}{cc}
\hline
$\vect{E}$ & $|E|$ \\
\hline
$\langle 0,0,Y \rangle$ & 2 \\
$\langle 0,0,N \rangle$ & 1 \\
$\langle 0,1,Y \rangle$ & 3 \\
$\langle 1,0,N \rangle$ & 1 \\
$\langle 1,1,Y \rangle$ & 1 \\
\hline
\end{tabular}
\end{center}
\end{table}
\end{example}

Let $S = (U, A)$ be a decision table, two attribute sets $P$ and $Q$, $P \subseteq Q \subseteq A$. We define partial relations $\preceq, \succeq$ as follows: 
$\vect E_P \preceq \vect E_Q$ (or $\vect E_Q \succeq \vect E_P$) if and only if $\forall p_i \in P$, there exists $q_j = \{p_i\}\cap Q$ such that $v_{q_j} = v_{p_i}$, where $P=\{p_1,\cdots, p_{|P|}\}$, $Q=\{q_1,\cdots, q_{|Q|}\}$, $\vect E_P=\langle v_{p_1}, v_{p_2}, \cdots, v_{p_{|P|}}\rangle$ and $\vect E_Q=\langle v_{q_1},v_{q_2},\cdots, v_{{q_{|Q|}}}\rangle$. 
With these notation, we define the granulating relation as follows. 
\begin{definition}[Granulating Relation] 
Let $G^{(P)}$ and $G^{(Q)}$ be two granularity representations of the decision table $S=(U, A)$ w.r.t. $P$ and $Q$, $P \subseteq Q \subseteq A$. 
$G^{(P)} \sqsubseteq G^{(Q)}$ (or $G^{(P)} \sqsupseteq G^{(Q)}$), if and only if 
$\forall E_P \in U/P$, $\exists E_Q \in U/Q$ such that $\vect E_P \preceq \vect E_Q$ and $E_Q \subseteq E_P$. 
\end{definition}
The granulating relation reveals the relationships between different granularity representations. Subsequently, we introduce two operations, coarsing and refining, that formally describes how to switch between two granularity representations. 

\begin{cor}[Coarsing]\label{cor:coarsing}
Given a granularity representation $G^{(Q)}=\{(\vect E_Q, |E_Q|) : E_Q\in U/Q\}$, $\forall P \subseteq Q$, the coarse granularity representation $G^{(P)}=\{(\vect E_P, |E_P|) : E_P\in U/P\}$ is computed by
\begin{equation}
\forall E_P \in U/P, E_P = \cup \{E_Q\in U/Q | \vect E_P \preceq \vect E_Q \} \nonumber
\end{equation}
\end{cor}

\begin{cor}[Refining]\label{cor:fining}
Given a granularity representation $G^{(P)}=\{(\vect E_P, |E_P|) : E_P\in U/P\}$, $\forall Q \supseteq P$, the refining granularity representation $G^{(Q)}=\{(\vect E_Q, |E_Q|) : E_Q\in U/Q\}$ is computed by
\begin{equation}
G^{(Q)}=\{(\vect E_Q, |E_Q|) : E_Q \in \{E_P/Q-P|E_P\in U/P\} \} \nonumber
\end{equation}
\end{cor}

\begin{figure}[!htbp]
\begin{center}
\includegraphics*[width=0.45\textwidth]{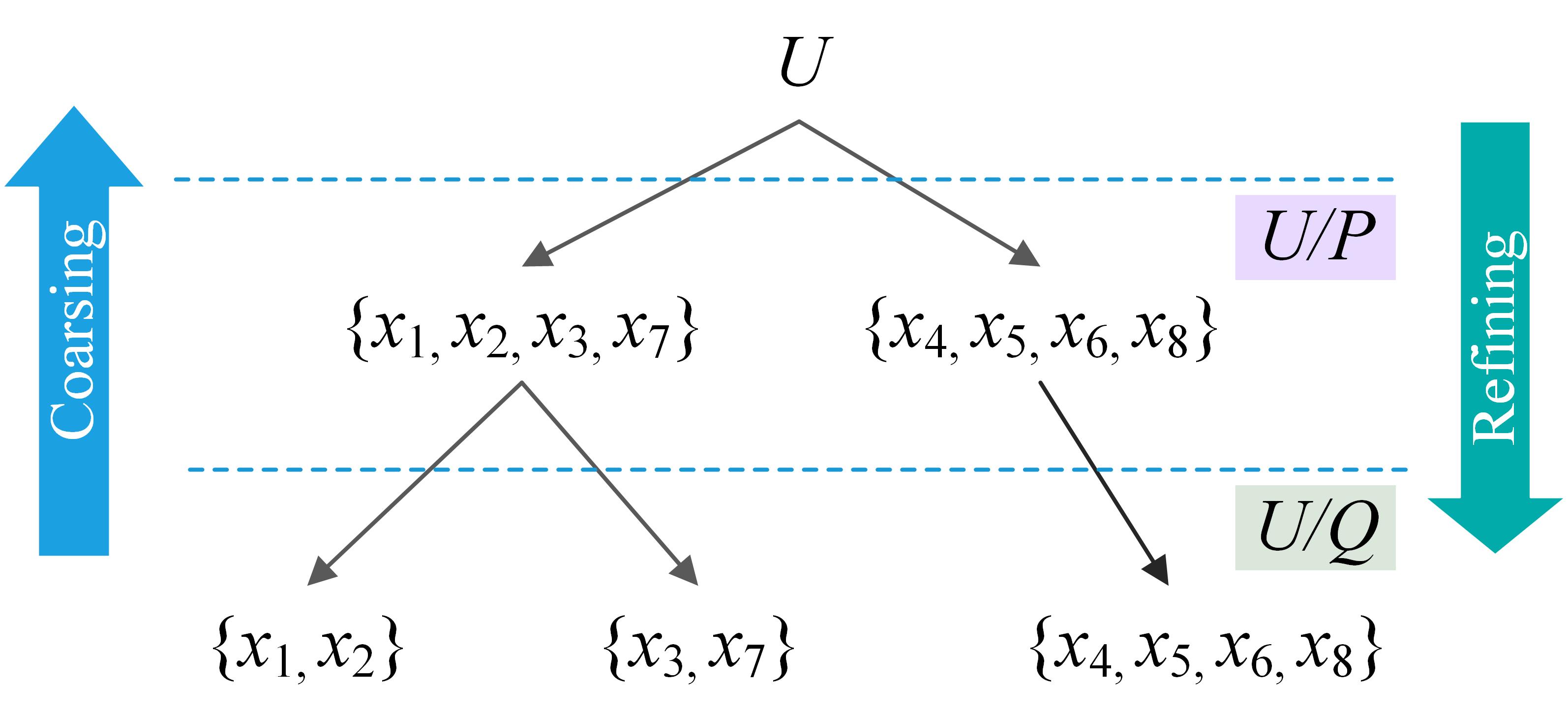}
\end{center}
\vspace*{-10pt}
\caption{Coarsing \& Refining between granularity representations} \label{fig:exampleGrC1}
\end{figure}

\begin{example}[Example 1 continued]
Let $P=\{a_2\}$ and $Q=\{a_2, D\}$. So we have $U/P=\{\{x_1,x_2,x_3,x_7\}, \{x_4,x_5,x_6,x_8\}\}$ and $U/Q=\{\{x_1,x_2\},\{x_3,x_7\},\{x_4,x_5,x_6,x_8\}\}$. Figure~\ref{fig:exampleGrC1} depicts the coarsing and refining operations. 
\end{example}

\section{Implementation}\label{sec:Im}
\begin{algorithm*}[!ht]
\KwIn{Input files: $input$ (\textit{i.e.}, decision table $S=(U, C\cup D)$), attribute importance measure metric $\Delta$, threshold $\epsilon$}
\KwOut{An attribute reduct $\mathcal R$}
\tcp{Initialization: constructing a granularity representation $G^{(C\cup D)}$, namely $aRDD$ }

$aRDD \defeq G^{(C\cup D)} \longleftarrow $ sc.\textFile{$input$}\\
\nonl \qquad\qquad\qquad\qquad\qquad\quad\   .\map{$line \Rightarrow$ parseVector()}\\
\nonl \qquad\qquad\qquad\qquad\qquad\quad\   .\reduceByKey{add}\;
$aRDD$.\cache{} \;
  $Cands \longleftarrow \{C\} \cup \{C\setminus\{a\}| a \in C\}$\;
  \ForPar(\tcp*[h]{Model parallelism: multiprocessing execution}){$B \in Cands$}{
    $\Theta(D|B)=$ \textsc{computingEF}$(aRDD,D,B,\Delta)$\;
  }
  \For{$a \in C$}{
    $Sig^{inner}_\Delta(a,C,D) \longleftarrow \Theta(D|C\setminus\{a\}) - \Theta(D|C)$\;
  }

  $Core=\{a| Sig^{inner}_\Delta(a,C,D)>\epsilon, a\in C\}$\;
  $\mathcal R \longleftarrow Core$\;
  \While{stopping criterion not met $\&$ $C\setminus\mathcal R \ne \emptyset$}{
    \ForPar(\tcp*[h]{Model parallelism: multiprocessing execution}){$a \in C\setminus \mathcal R$}{
      $\Theta(D|\mathcal R\cup \{a\})=$ \textsc{computingEF}$(aRDD, D, \mathcal R \cup \{a\},\Delta)$\;
    }
    $a_{opt}=\argmin\limits_{a\in C\setminus\mathcal R}\{\Theta(D|\mathcal R\cup \{a\})\}$\;
    $\mathcal R \longleftarrow \mathcal R \cup \{a_{opt}\}$\;
  }
  \Return $\mathcal R$
\\
\tcp{Computing Evaluation Function's Value}
\SetKwProg{myproc}{Procedure}{}{}
\SetKwFunction{EF}{\textsc{\texttt{computingEF}}}
\nonl \myproc{\EF{$aRDD, D, B, \Delta$}} 
{
$bRDD \defeq G^{(B \cup D)} \longleftarrow aRDD$.\map{$\left(\vect E_{C\cup D}, |E_{C\cup D}|\right) \Rightarrow \left(\vect E_{B\cup D}, |E_{B\cup D}|\right)$}\\
$\Theta(D|B)=bRDD$.\map{$\left(\vect E_{B\cup D}, |E_{B\cup D}|\right) \Rightarrow \left(\vect E_B, (\vect E_D, |E_{B\cup D}|)\right)$}\\
\nonl \qquad\qquad\qquad\qquad\ .\reduceByKey{$S_i \Rightarrow$ $\theta (S_i)$}\\
\nonl \qquad\qquad\qquad\qquad\  .\Sum{}\;
\Return $\Theta(D|B)$
}
\caption{PLAR}\label{alg:plar} 
\end{algorithm*}

In this section, we describe the implementation of our algorithms on top of \Spark{} \cite{Spark}. 
As introduced in Section~\ref{sec:Pa}, the key components of the MDP framework are model-parallelism and data-parallelism. 
To achieve the best possible performance, we consider both GrC-based initialization and MDP. 
Algorithm~\ref{alg:plar} outlines the details of the parallel large-scale attribute reduction (PLAR) algorithm. 

\begin{itemize}
\item The first stage leverages the granularity representation introduced in Section~\ref{sec:GrC} to do the GrC-based initialization (lines 1-2), which is consisted of 1) loading data from distributed file system (\textit{e.g.} HDFS) via the function ``sc.\textFile{}''; 2) constructing the granularity representation of the decision table;  3) caching the granularity representation in the distributed memory (line 2). It means that loading data and constructing the granularity representation are only executed once in the whole computing processing.  
\item The second stage is getting the attribute core (lines 3-8). We first generates the feature candidates (line 3), then use MP to execute a parallel-for loop to compute evaluation function's value simultaneously (lines 4-5) where \textsc{computingEF} is a DP-based method for computing evaluation function's value. After that, we compute the significance of all candidates (lines 6-7), and select the satisfied attribute (\textit{i.e.} greater than a threshold here) into the attribute core (line 8). 
\item The third stage is computing the attribute reduct (lines 9-14). First, we initialize the attribute reduct $\mathcal R$ as $Core$ (line 9). 
We invoke the iterative processing of feature selection which  selects the optimal feature into the attribute reduct $\mathcal R$ (lines 10-14). In each iteration, we compute evaluation function's value using MP simultaneously (lines 11-12). Then we select the best attribute (line 13) and put it into $\mathcal R$ (line 14).  
\item Procedure \textsc{computingEF} describes the processing for computing evaluation function's value using the \Spark{}-style language. 
Given the granularity representation $aRDD \defeq G^{C\cup D}$, we first compute the granularity representation $G^{B\cup D}$ according to Corollary~\ref{cor:coarsing} (line 16). Then, we use DP to compute the evaluation function's value (line 17) and the details are already described in Section~\ref{sec:EF}. 
To easily understand it, we give an illustration as follows.  
\end{itemize}

\begin{figure}[!htbp]
\begin{center}
\includegraphics*[width=0.45\textwidth]{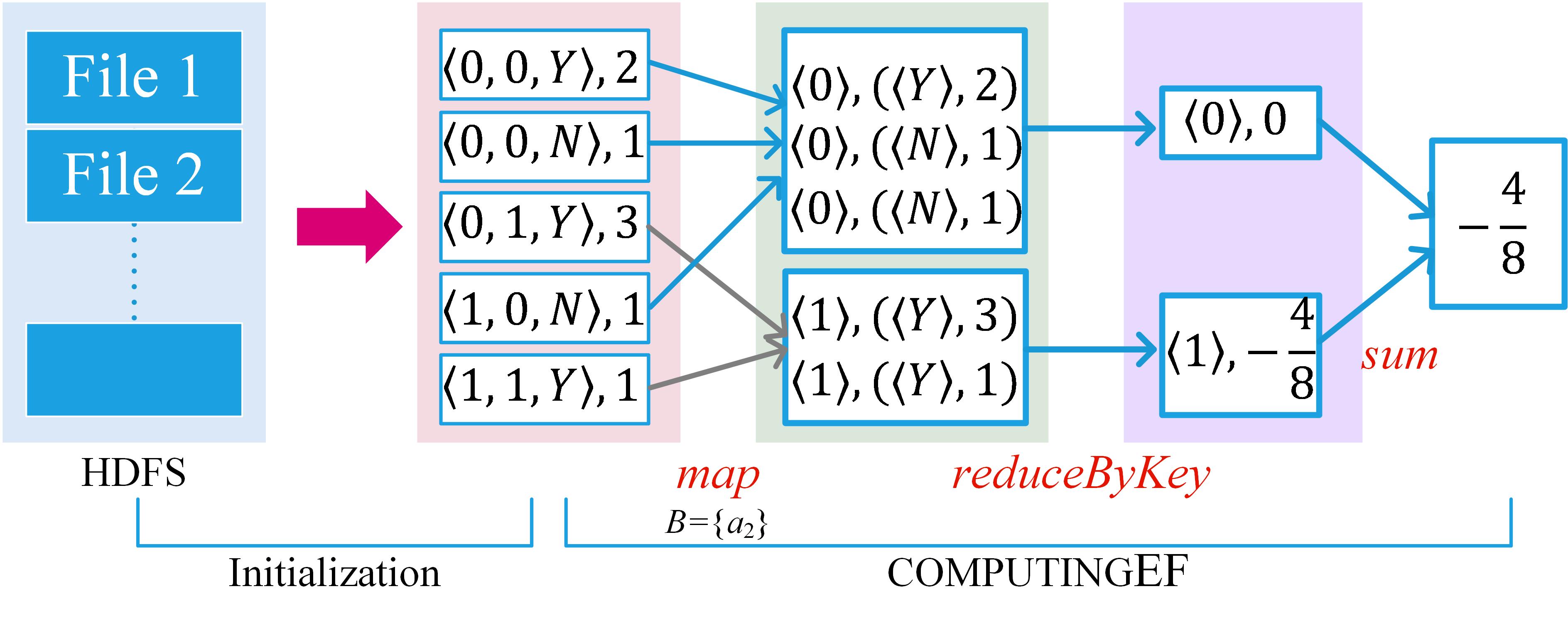}
\end{center}
\vspace*{-10pt}
\caption{Example for computing evluation function's value on \Spark{}} \label{fig:example2}
\end{figure}

\begin{example}[Example 1 continued, $\Delta = PR$]
Figure~\ref{fig:example2} shows an example for computing evluation function's value on \Spark{}. \\
(1) The original decision table is stored in HDFS. 
With GrC initialization, we can get the granularity representation of the decision table. \\
(2) Here, we suppose to evaluate $B=\{a_2\}$. Using \map{}, it produces $\left(\vect E_B, (\vect E_D, |E_{B\cup D}|)\right)$. Taking ``$\langle 0,0,Y \rangle , 2$'' for example, we have $\vect E_B=\vect E_{a_2} = \langle 0 \rangle$, $\vect E_D = \langle Y\rangle$ and $|E_{B\cup D}| = 2$. Therefore,  its output is ``$\langle 0 \rangle, (\langle Y\rangle, 2)$''. So are others. \\
(3) \reduceByKey{} is employed to aggregate the values with same key and then compute the sub-evaluation function's value. Taking ``$key = \langle 1 \rangle$'' for example, according to PR's definition (see Table~\ref{tab:4Methods:Par}), $-\frac{|E_i|sgn_{PR}(E_i)}{|U|} = -\frac{4}{8}$. 
\\ 
(4) \Sum{} is computing the sum of all sub-evaluation functions' value. 
\end{example}

\subsection{Complexity Analysis}
Assume that $S=(U, A)$ is a decision table, therefore it requires $\mathcal O(|U||A|)$ to store the data. With the GrC-based initialization, the data is only read once when loading, and only the feature vectors of equivalence classes and associated cardinality are stored. Thus, the space complexity is 
\begin{equation}
\mathcal O(|U/A|(|A|+1))  = \mathcal O(|U/A||A|)\nonumber
\end{equation}
where $|U/A|$ is the total number of equivalence classes over $U$ w.r.t. $A$. 


\input{expt}
\input{plar_draft.bbl.tex}

\end{document}

%% file: expt.tex
\section{Experimental Results}\label{sec:E}
In this section, we evaluate PLAR experimentally. 
We first describe the experimental setup in Section~\ref{sec:setup}. 
Then we report the results in terms of effectiveness and efficiency in Section~\ref{sec:effectiveness} and Section~\ref{sec:efficiency}, respectively. 
We show effect of PLAR's components in Section~\ref{sec:effect}. 

\subsection{Setup}\label{sec:setup}

\noindent\textbf{Cluster.} We perform all experiments on a cluster of 19 machines, each with at least 8-core 2.0 GHz processors, more than 8 GB RAM, running Cent OS 6.5. 
All machines are connected via a Gigabit network. We deploy \Spark{} of version 1.x on each machine in the cluster, and configure one machine as the master node and the other ones as slaves. The total number of cores used by \Spark{} is 128.  
All our algorithms are implemented in the Python programming language and PySpark. 

\noindent\textbf{Datasets.} We tested on 13 benchmark datasets of various scales. The description of each dataset, including its samples and features, is shown in Table~\ref{tab:plar:data}. Among them, Datasets 1-9 are small datasets, downloaded from UCI dataset repository\footnote{\url{http://archive.ics.uci.edu/ml/}}; 
Datasets 10 and 11 have millions of samples, and are respectively obtained from KDD CUP 1999\footnote{\url{http://kdd.ics.uci.edu/databases/kddcup99/kddcup99.html}} and generated by the WEKA\footnote{\url{http://www.cs.waikato.ac.nz/ml/weka/}} data generator. 
Dataset 12 is a high-dimensional dataset, which was selected as the NIPS 2003 feature selection challenge\footnote{\url{https://archive.ics.uci.edu/ml/datasets/Gisette}}; 
Dataset 13 is really big dataset (large \& high-dimension), collected from the astronomical data repository\footnote{\url{http://www.sdss.org/data/}}. 
All datasets are uploaded into the HDFS before computation. 

\begin{table}[!htbp]
\tabcolsep 0pt \caption{Datasets used in the experiments}\label{tab:plar:data}  \label{tab:vsQianAI} \label{tab:vsHadoopMR} \vspace*{-20pt}
    \begin{flushleft}
    \def\temptablewidth{0.48\textwidth}
        {\rule{\temptablewidth}{1pt}}
        \begin{tabular*}{\temptablewidth}{@{\extracolsep{\fill}}llrrrc}
            {No.} & Dataset & Samples & Features & Classes & Note \\
            \hline
  1 & Mushroom & 5644 & 22 & 2 & \multirow{9}*{small}\\
  2 & Tic-tac-toe & 958 & 9 & 2\\
  3 & Dermatology & 358 & 34 & 6\\
  4 & Kr-vs-kp & 3196 & 36 & 2\\
  5 & Breast-cancer-wisconsin & 683 & 9 & 2\\
  6 & Backup-large.test  & 376 & 35 & 19\\
  7 & Shuttle & 58000 & 9 & 7\\
  8 & Letter-recognition & 20000 & 16 & 26\\
  9 & Ticdata2000 & 5822 & 85 & 2\\
  \hline
  10 & KDD99 & 5,000,000 & 41 & 23 & \multirow{2}*{large}\\
  11 & WEKA15360 & 15,360,000 & 20 & 10\\
  \hline
  \multirow{2}*{12} &  \multirow{2}*{Gisette} &  \multirow{2}*{6000} &  \multirow{2}*{5000} &  \multirow{2}*{2} & high- \\
  &&&&& dimension \\
  \hline
  \multirow{2}*{13} &  \multirow{2}*{SDSS} &  \multirow{2}*{320,000} &  \multirow{2}*{5201} &  \multirow{2}*{17} & large \& high- \\
  &&&&& dimension \\
        \end{tabular*}
        {\rule{\temptablewidth}{1pt}}
        \end{flushleft}
\end{table}

\noindent\textbf{Parameter.} The \textit{model parallelism level} is a parameter in PLAR, which means the maximum concurrency number for computing attributes'  significance measures.  

\subsection{Effectiveness}\label{sec:effectiveness}
To evaluate the effectiveness of PLAR, we compare it with two baselines, including HAR and FSPA, which are described below.
\begin{itemize}
\item HAR is an original forward Heuristic Attribute Reduction algorithm (see Algorithm~1). 
\item FSPA \cite{Qian2010AI} is a general Feature Selection based on the Positive Approximation, which is a state-of-the-art algorithm in a single machine. 
\end{itemize}
We evaluate PLAR by four representative significance measures of attributes, including PR, SCE, LCE and CCE (see details in Section~\ref{sec:sub:EF}). 
Since computing and storing for the traditional single-machine-based approaches is impractical for large datasets, we use Datasets 1-9 of Table~\ref{tab:plar:data} for this test. 
The used cores' number and the model parallelism level are both set as 8 for PLAR.  

Tables~\ref{tab:vsQianAI:PR}-\ref{tab:vsQianAI:CCE} show the elapsed time and attribute reduction of the algorithms HAR, FSPA and PLAR using four different attribute measures. 
Taking PR for example, from Table~\ref{tab:vsQianAI:PR}, we can observe that the feature subset obtained by PLAR is the same as the algorithms HAR and FSPA. The similar results in Tables~\ref{tab:vsQianAI:SCE}-\ref{tab:vsQianAI:CCE} demonstrate that our proposed parallel algorithm can produce the consistent attribute reduction. 
We find that FSPA is much faster in several small datasets such as Breast-cancer-wisconsin, and PLAR can achieve better performance in some large datasets, including Shuttle, Letter-recognition and Ticdata2000. 

\begin{table*}[!htbp]
\tabcolsep 0pt \caption{The time and attribute reduction of the algorithms HAR-PR, FSPA-PR and PLAR-PR}\label{tab:vsQianAI:PR} \vspace*{-15pt}
\begin{threeparttable}
    \begin{flushleft}
    \def\temptablewidth{\textwidth}
        {\rule{\temptablewidth}{1pt}}
        \begin{tabular*}{\temptablewidth}{@{\extracolsep{\fill}}lrlrlrlr}
            \multirow{2}*{Dataset} & \multirow{2}*{Original features} & \multicolumn{2}{c}{HAR-PR \cite{Qian2010AI}} & \multicolumn{2}{c}{FSPA-PR \cite{Qian2010AI}} & \multicolumn{2}{c}{PLAR-PR}\\
            \cline{3-4} \cline{5-6} \cline{7-8}
             & & Time (s)  & Selected features & Time (s) & Selected features &Time (s) & Selected features  \\
            \hline
Mushroom&22&24.875&3&20.453&3&5.698&3\\
Tic-toc-toe&9&0.359&8&0.313&8&4.650&8\\
Dermatology&34&0.844&10&0.438&10&10.420&10\\
Kr-vs-kp&36&28.031&29&21.578&29&5.632&29\\
Breast-cancer-wisconsin&9&0.125&4&0.094&4&3.465&4\\
Backup-large.test&35&0.656&10&0.422&10&9.621&10\\
Shuttle&9&906.063&4&712.250&4&3.964&4\\
Letter-recognition&16&282.641&11&112.625&11&7.938&11\\
Ticdata2000&85&886.453&24&296.375&24&55.963&24\\
        \end{tabular*}
        {\rule{\temptablewidth}{1pt}}
        \end{flushleft}
\end{threeparttable}
\vspace*{10pt}
\tabcolsep 0pt \caption{The time and attribute reduction of the algorithms HAR-SCE, FSPA-SCE and PLAR-SCE}\label{tab:vsQianAI:SCE} \vspace*{-15pt}
\begin{threeparttable}
    \begin{flushleft}
    \def\temptablewidth{\textwidth}
        {\rule{\temptablewidth}{1pt}}
        \begin{tabular*}{\temptablewidth}{@{\extracolsep{\fill}}lrlrlrlr}
            \multirow{2}*{Dataset} & \multirow{2}*{Original features}& \multicolumn{2}{c}{HAR-SCE \cite{Qian2010AI}} & \multicolumn{2}{c}{FSPA-SCE \cite{Qian2010AI}} & \multicolumn{2}{c}{PLAR-SCE} \\
            \cline{3-4} \cline{5-6} \cline{7-8}
             && Time (s) & Selected features & Time (s) & Selected features &Time (s) & Selected features\\
            \hline
Mushroom&22&162.641&4&159.594&4&6.575&4\\
Tic-toc-toe&9&4.500&8&3.109&8&4.771&8\\
Dermatology&34&5.313&11&1.984&11&13.451&11\\
Kr-vs-kp&36&149.625&29&105.984&29&5.547&29\\
Breast-cancer-wisconsin&9&1.344&4&0.844&4&3.667&4\\
Backup-large.test&35&4.359&10&1.766&10&10.894&10\\
Shuttle&9&12665.391&4&10153.172&4&4.065&4\\
Letter-recognition&16&7015.703&11&2740.250&11&7.953&11\\
Ticdata2000&85&8153.656&24&1043.891&24&53.953&24\\
        \end{tabular*}
        {\rule{\temptablewidth}{1pt}}
        \end{flushleft}
\end{threeparttable}
\vspace*{10pt}
\tabcolsep 0pt \caption{The time and attribute reduction of the algorithms HAR-LCE, FSPA-LCE and PLAR-LCE}\label{tab:vsQianAI:LCE} \vspace*{-15pt}
\begin{threeparttable}
    \begin{flushleft}
    \def\temptablewidth{\textwidth}
        {\rule{\temptablewidth}{1pt}}
        \begin{tabular*}{\temptablewidth}{@{\extracolsep{\fill}}lrlrlrlr}
            \multirow{2}*{Dataset} & \multirow{2}*{Original features}& \multicolumn{2}{c}{HAR-LCE \cite{Qian2010AI}} & \multicolumn{2}{c}{FSPA-LCE \cite{Qian2010AI}} & \multicolumn{2}{c}{PLAR-LCE} \\
           \cline{3-4} \cline{5-6} \cline{7-8}
            & & Time (s) & Selected features & Time (s) & Selected features &Time (s) & Selected features \\
            \hline
Mushroom&22&300.219&4&294.000&4&6.728&4\\
Tic-toc-toe&9&8.734&8&5.781&8&4.560&8\\
Dermatology&34&10.453&10&3.750&10&13.073&10\\
Kr-vs-kp&36&1156.125&29&191.125&29&5.275&29\\
Breast-cancer-wisconsin&9&3.125&5&1.672&5&3.642&5\\
Backup-large.test&35&9.844&10&3.219&10&10.016&10\\
Shuttle&9&24883.625&4&20228.391&4&3.888&4\\
Letter-recognition&16&15176.766&12&5558.781&12&7.767&12\\
Ticdata2000&85&27962.625&24&1805.563&24&54.684&24\\
        \end{tabular*}
        {\rule{\temptablewidth}{1pt}}
        \end{flushleft}
\end{threeparttable}
\vspace*{10pt}
\tabcolsep 0pt \caption{The time and attribute reduction of the algorithms HAR-CCE, FSPA-CCE and PLAR-CCE}\label{tab:vsQianAI:CCE} \vspace*{-15pt}
\begin{threeparttable}
    \begin{flushleft}
    \def\temptablewidth{\textwidth}
        {\rule{\temptablewidth}{1pt}}
        \begin{tabular*}{\temptablewidth}{@{\extracolsep{\fill}}lrlrlrlr}
            \multirow{2}*{Dataset} & \multirow{2}*{Original features}& \multicolumn{2}{c}{HAR-CCE \cite{Qian2010AI}} & \multicolumn{2}{c}{FSPA-CCE \cite{Qian2010AI}} & \multicolumn{2}{c}{PLAR-CCE} \\
            \cline{3-4} \cline{5-6} \cline{7-8}
             && Time & Selected features & Time (s) & Selected features &Time (s) & Selected features \\
            \hline
Mushroom&22&166.922&4&159.641&4&3.278&4\\
Tic-toc-toe&9&6.766&8&3.141&88&4.822&8\\
Dermatology&34&5.828&10&2.266&10&11.433&10\\
Kr-vs-kp&36&149.750&29&105.750&29&12.470&29\\
Breast-cancer-wisconsin&9&1.359&4&0.891&4&2.323&4\\
Backup-large.test&35&4.578&9&1.984&9&3.373&9\\
Shuttle&9&13718.875&4&10948.922&4&3.850&4\\
Letter-recognition&16&7118.266&11&2610.359&11&8.110&11\\
Ticdata2000&85&8262.047&24&1048.578&24&59.114&24\\
        \end{tabular*}
        {\rule{\temptablewidth}{1pt}}
        \end{flushleft}
\end{threeparttable}
\end{table*}

\subsection{Efficiency}\label{sec:efficiency}
To measure the efficiency of PLAR, we evaluate it on four aspects below. 
\begin{itemize}
\item Comparison with single-machine algorithms: see Section~\ref{sec:small}. 
\item Comparison with distributed algorithms: see Section~\ref{sec:large}.  
\item Effect of model parallelism level: see Section~\ref{sec:high}. 
\item Speedup on large \& high-dimensional data: see Section~\ref{sec:large-high}.  
\end{itemize}

\subsubsection{Comparison with single-machine algorithms}\label{sec:small}
\begin{figure*}[!htbp]
\centering
\subfigure[Attribute Reduction Algorithms with PR]{\includegraphics[width=.9\textwidth]{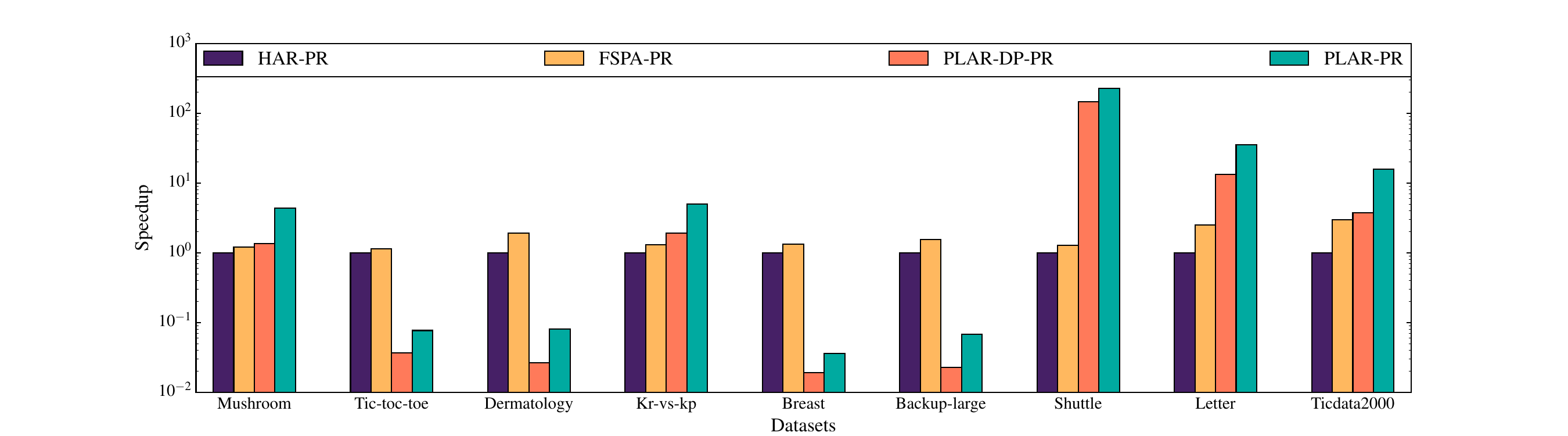}}
\subfigure[Attribute Reduction Algorithms with SCE]{\includegraphics[width=.9\textwidth]{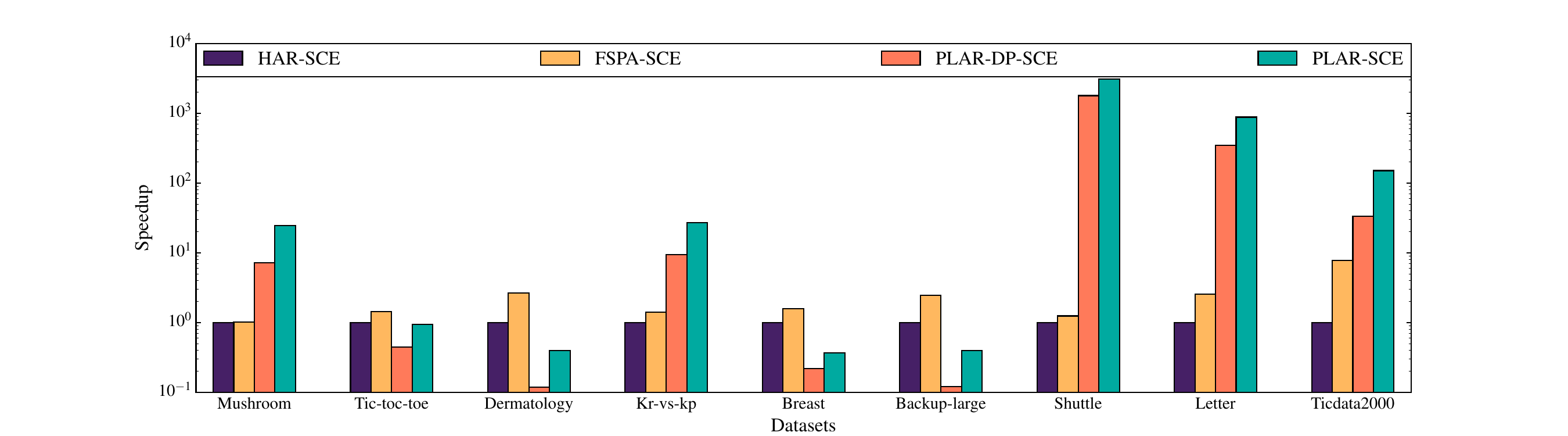}}
\subfigure[Attribute Reduction Algorithms with LCE]{\includegraphics[width=.9\textwidth]{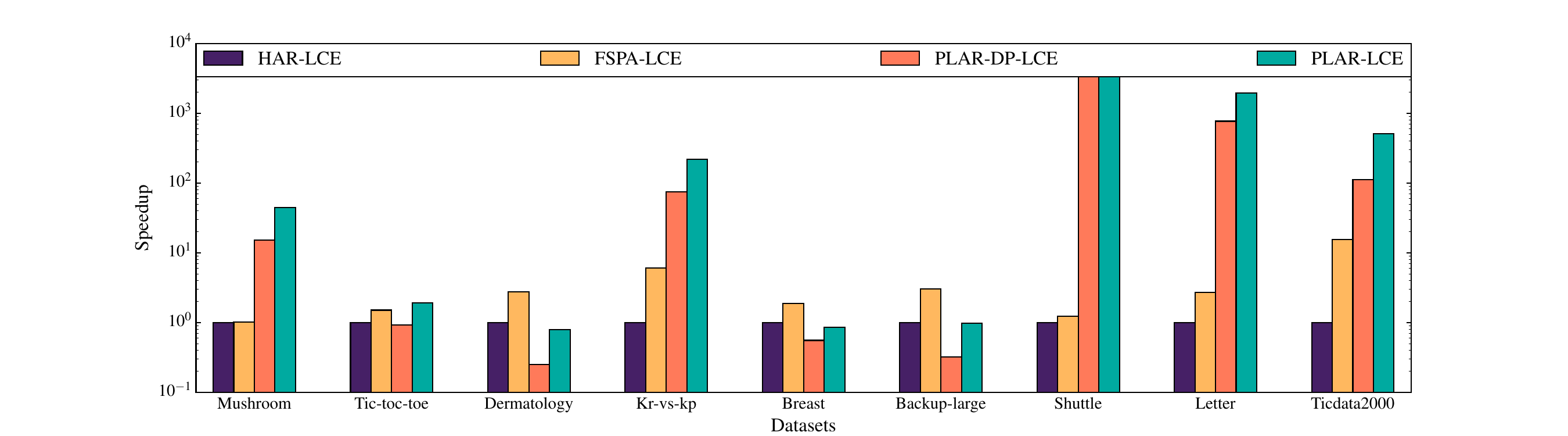}}
\subfigure[Attribute Reduction Algorithms with CCE]{\includegraphics[width=.9\textwidth]{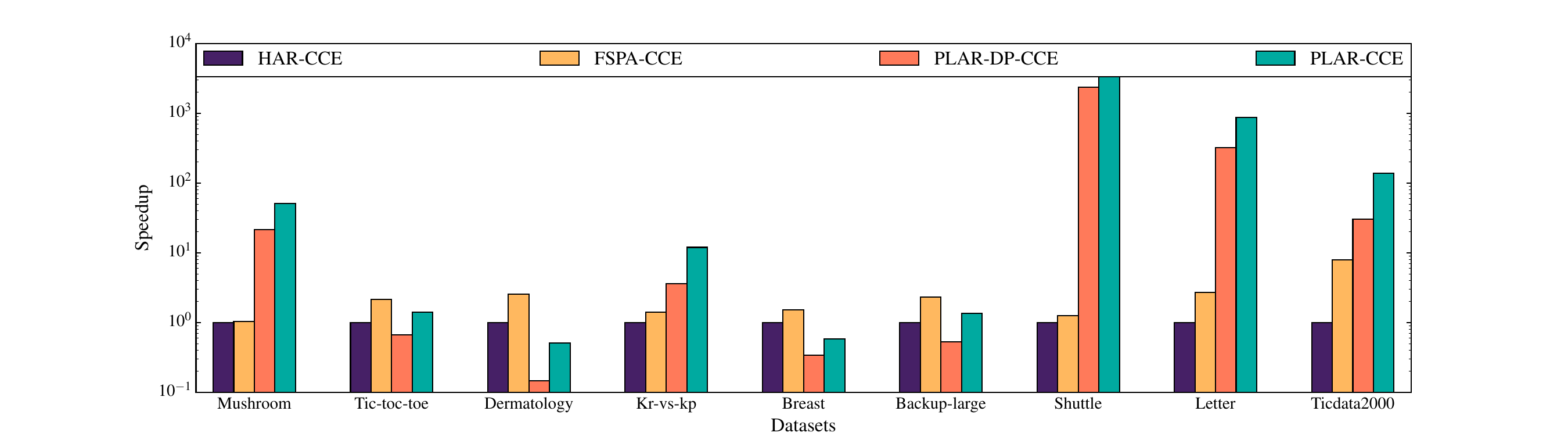}}
\\
\emph{Breast: Breast-cancer-wisconsin; Backup-large: Backup-large.test; Letter: Letter-recognition}
\caption{Speedup on small sample size datasets}
\label{fig:PLAR:vsQianAI}
\end{figure*}

In this section, we compare PLAR with the state-of-the-art single machine algorithms on 9 small datasets. The used datasets and compared baseline algorithms are same with ones in Section~\ref{sec:effectiveness}. Besides, we perform an experiment on an algorithm called PLAR-DP, which is a simplified version of PLAR that only utilizes the data parallelism without the model parallelism. 
Because the algorithms HAR and FSPA can only run in a single machine, we perform PLAR and PLAR-DP on a 8-core single machine as well. PLAR's model parallelism level is set as 8. 
We here use $\rm{speedup}=\frac{\texttt{running time of a certain algorithm}}{\texttt{running time of HAR}}$ to measure all algorithms. 
From Figure~\ref{fig:PLAR:vsQianAI},  we can see FSPA outperform other algorithms in the datasets Tic-tac-toe, Dermatology, Breast-cancer-wisconsin and Backup-large.test, whose samples are very few, all less than 1,000.  
As the sample size increases, PLAR performs better and better, especially in the datasets Shuttle and Letter-recognition whose samples are more than 10,000, where PLAR can achieve 100$\times$, even 1000$\times$ speedup. Taking LCE for example, against HAR, PLAR is about 6400$\times$ faster. 
The experimental results show that PLAR is always faster than PLAR-DP which verifies the benefit of the model parallelism. 

\subsubsection{Comparison with distributed algorithms}\label{sec:large}
\begin{figure}[!htbp]
\centering
\subfigure[Attribute Reduction Algorithms with PR]{\includegraphics[width=.48\textwidth]{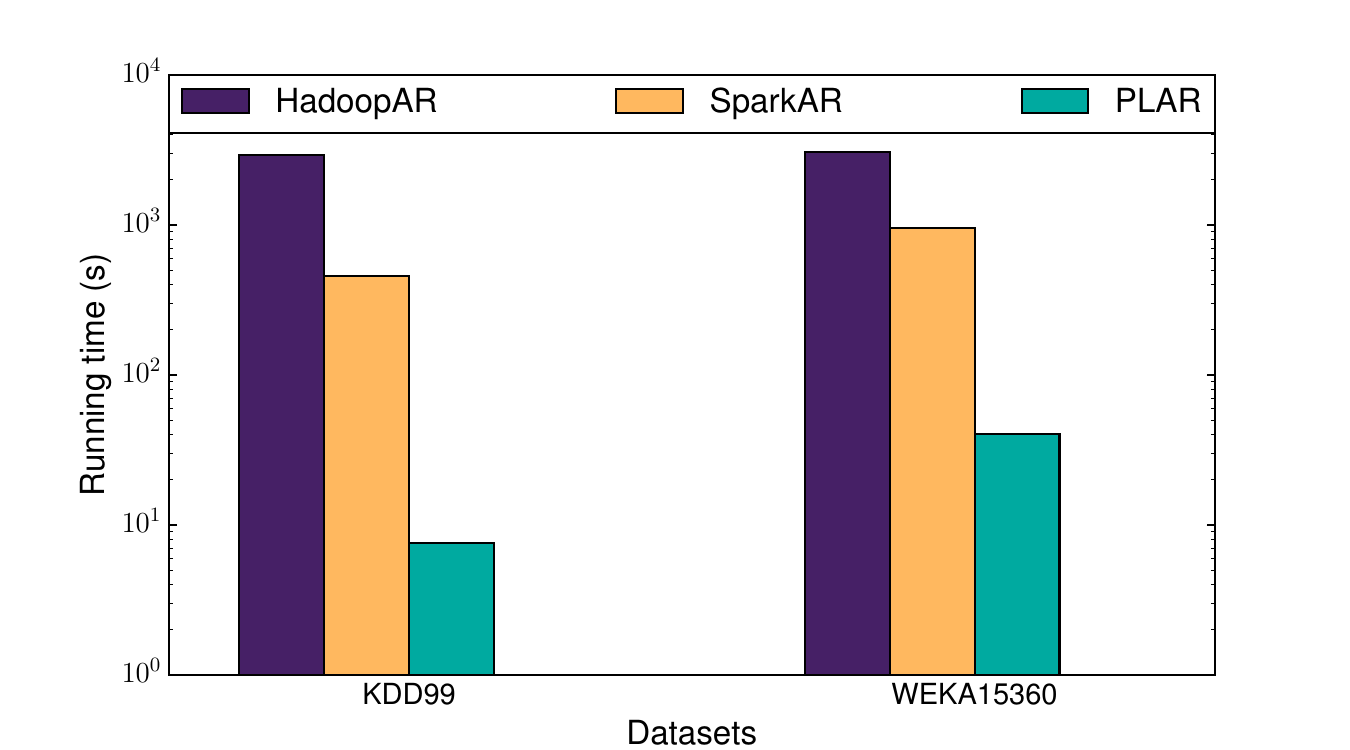}}
\subfigure[Attribute Reduction Algorithms with SCE]{\includegraphics[width=.48\textwidth]{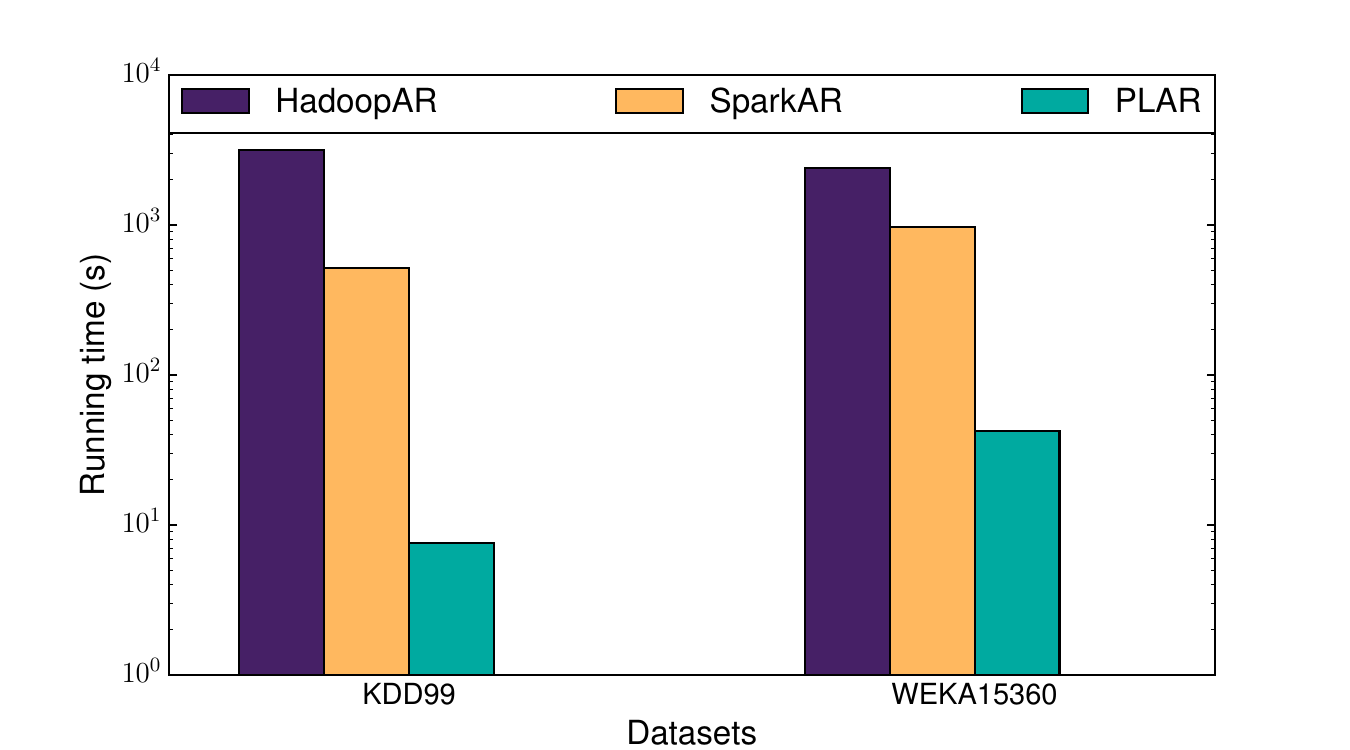}}
\subfigure[Attribute Reduction Algorithms with LCE]{\includegraphics[width=.48\textwidth]{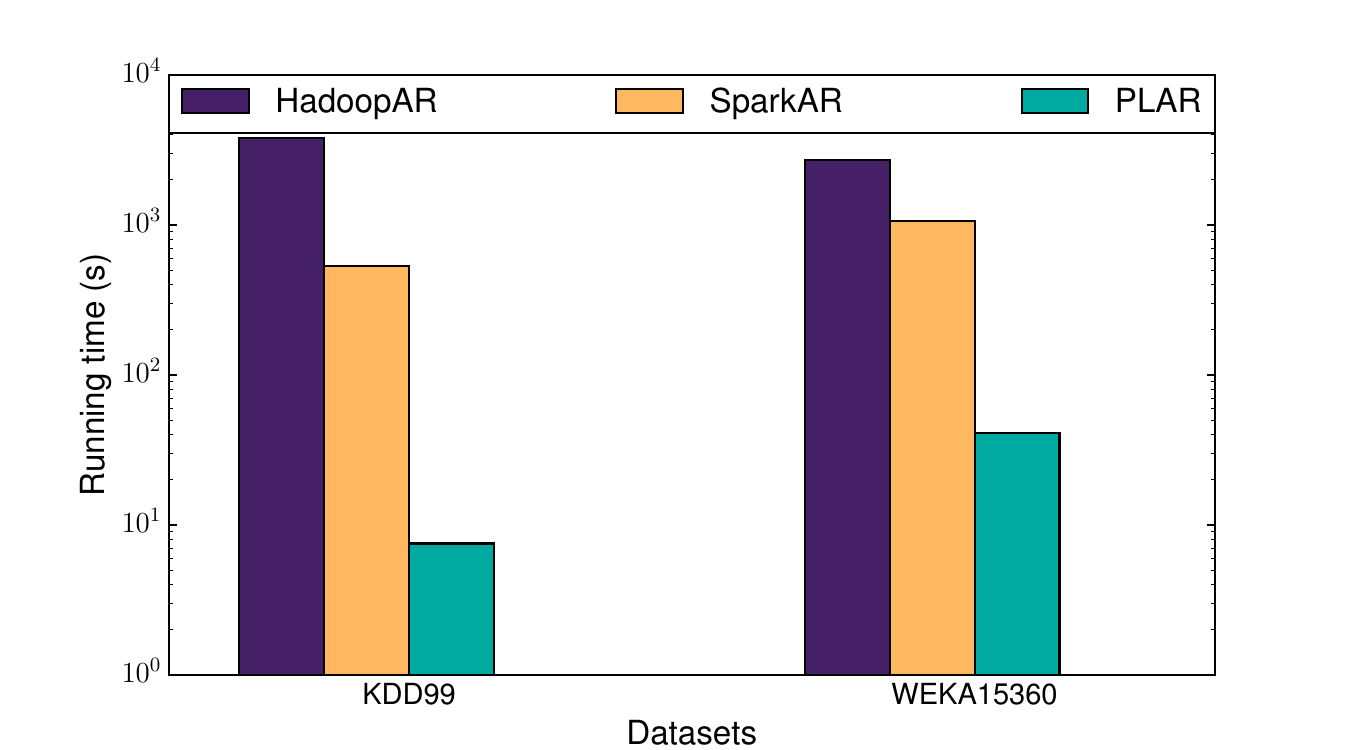}}
\subfigure[Attribute Reduction Algorithms with CCE]{\includegraphics[width=.48\textwidth]{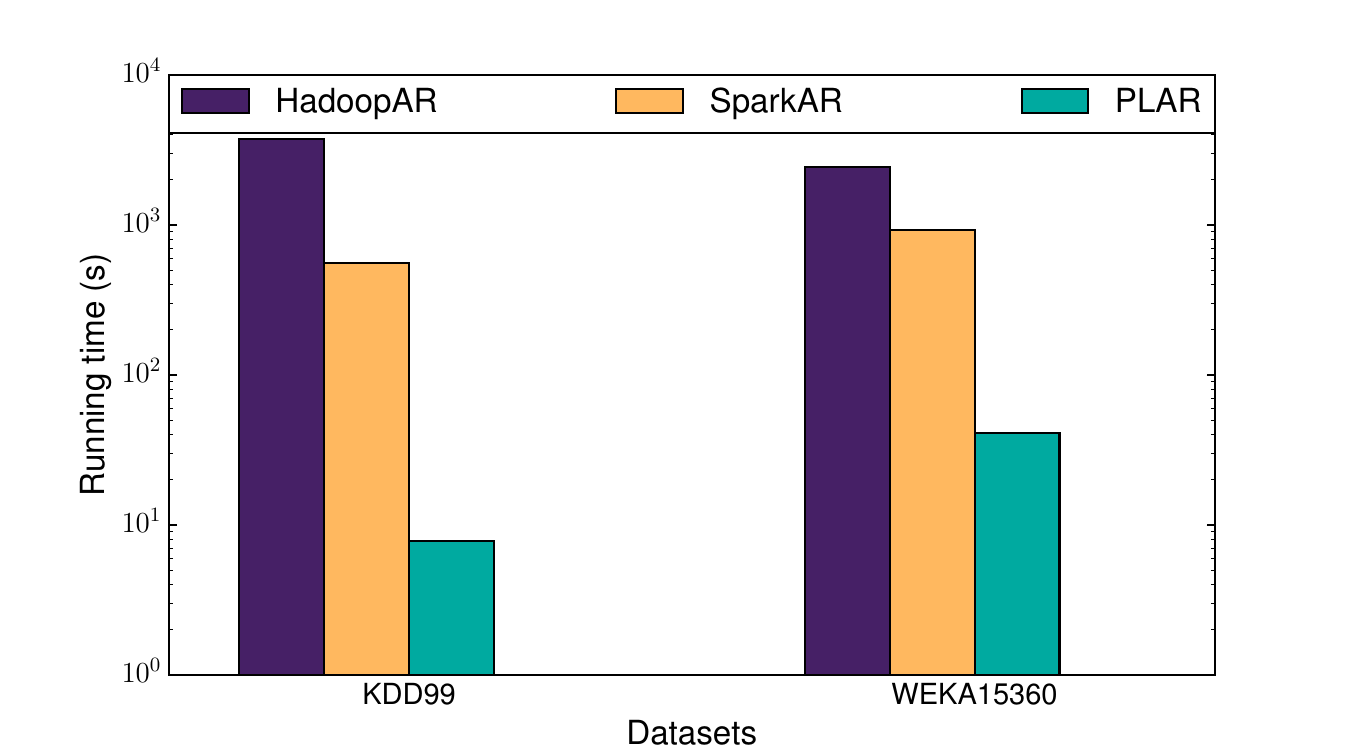}}
\footnotesize
\caption{The running time on different distributed algorithms}
\label{fig:PLAR:vsHadoopMR}
\end{figure}

In this experiment, we compare PLAR with some distributed algorithms, including HadoopAR, SparkAR and PLAR-DP, which are described as follows. 
\begin{itemize}
\item HadoopAR \cite{Zhang2013} is a parallel attribute reduction algorithm which is implemented in the Hadoop\footnote{\url{http://hadoop.apache.org/}} platform. 
\item SparkAR is a modified version of HadoopAR which is implemented on the Spark platform. 
\end{itemize}
In all of these experiments, we set the same core number, \textit{i.e.}, 16. 
Figure~\ref{fig:PLAR:vsHadoopMR} shows the performance comparison of these 4 distributed algorithms. 
It is easy to see that SparkAR is much faster than HadoopAR because HadoopAR has to read the data from HDFS each time when evaluating the attribute. On the contrary, Spark-based algorithms always read the data once into the distributed memory which makes the same parallel method can achieve up to 100$\times$ speedup on Spark than on Hadoop. PLAR always outperforms other algorithms.  

To display the results intuitively, $\rm{speedup}=\frac{\texttt{running time of a certain algorithm}}{\texttt{running time of HadoopAR}}$ is employed here, as shown in Table~\ref{fig:PLAR:vsHadoopMR}. It demonstrates that SparkAR is always $6-7\times$ faster than HadoopAR in the dataset KDD99, and reaches $2.48-3.24\times$ in the dataset WEKA15360. PLAR-DP and PLAR perform outstandingly, and can speed up more than 200 times in KDD99. Specifically, when using the attribute measure LCE, PLAR achieves 500-fold improvement. 
In WEKA15360, PLAR is 50-fold faster than HadoopAR over all four attribute measures, and obtains $75.88\times$ speedup when using the attribute measure PR. 
 
\begin{table}[!htbp]
\tabcolsep 0pt \caption{Speedup of different distributed algorithms}\label{tab:vsHadoopMR_speedup} \vspace*{-15pt}
    \begin{flushleft}
    \def\temptablewidth{0.48\textwidth}
        {\rule{\temptablewidth}{1pt}}
        \begin{tabular*}{\temptablewidth}{@{\extracolsep{\fill}}llrrr}
            Attribute  & \multirow{2}*{Dataset} & \multirow{2}*{HadoopAR} & \multirow{2}*{SparkAR} & \multirow{2}*{PLAR} \\
            measure & \\
            \hline
\multirow{2}*{PR}
&KDD99&1.00&6.39&381.89\\
&WEKA15360&1.00&3.24&75.88\\
\hline
\multirow{2}*{SCE}
&KDD99&1.00&6.08&414.85\\
&WEKA15360&1.00&2.48&57.11\\
\hline
\multirow{2}*{LCE}
&KDD99&1.00&7.14&502.87\\
&WEKA15360&1.00&2.55&65.71\\
\hline
\multirow{2}*{CCE}
&KDD99&1.00&6.75&480.16\\
&WEKA15360&1.00&2.63&59.57\\
        \end{tabular*}
        {\rule{\temptablewidth}{1pt}}
        \end{flushleft}
\end{table}

\subsubsection{Speedup on large \& high-dimensional data}\label{sec:large-high}
In this experiment, we test our algorithms with different numbers of cores for attribute reduction on the dataset SDSS of Table~\ref{tab:plar:data}. 
Because the overall running time for this dataset is very large, we test two configurations: 32 cores and 128 cores. 
Taking SCE and 128 cores for example, we record the running time for each iteration, the first 5 iterations cost 7312, 6696, 6793, 7659 and 7035 seconds, respectively. For the first iteration, there are totally 5201 feature candidates. Therefore, the average elapsed time for evaluating one attribute is about $1.406$ seconds. 
When the same test is ran on 32 cores, the first iteration takes 24180 seconds that means $4.649$ seconds per attribute. 
Hence, 4 times cores can achieve $\frac{4.649}{1.406} \approx 3.3 \times$ speedup. 

We do the similar experiments on all four attribute measure methods, shown in Table~\ref{tab:SDSS_running_time}. 
Comparing with 32 cores, performance of PR, SCE, LCE and CCE on 128 cores is $3.27$, $3.31$, $3.38$ and $3.29$ times faster, respectively. 
It demonstrates that it can efficiently reduce the running time as the core number increases. 
\begin{table}[!htbp]
\tabcolsep 0pt \caption{The running time for one iteration on SDSS (unit: second)}\label{tab:SDSS_running_time} \vspace*{-20pt}
    \begin{flushleft}
    \centering
    \def\temptablewidth{0.48\textwidth}
        {\rule{\temptablewidth}{1pt}}
        \begin{tabular*}{\temptablewidth}{@{\extracolsep{\fill}}ccc}
        Attribute measure & 128 cores & 32 cores \\
\hline
PR&7432&24274\\
SCE&7312&24181\\
LCE&7207&24372\\
CCE&7383&24295\\
\end{tabular*}
        {\rule{\temptablewidth}{1pt}}
        \end{flushleft}
\end{table}

\subsection{Effect of components of PLAR}\label{sec:effect}
The key components of PLAR are GrC-based initialization, data-parallelism, and model-parallelism. 
The effect of data-parallelism has been richly verified in Section~\ref{sec:efficiency}. 
Here, we mainly shows the effects of the GrC-based initialization and model-parallelism. 

\subsubsection{Effect of GrC-based initialization}
\begin{figure}[!htbp] 
\centering
\subfigure[PR]{\includegraphics[width=.24\textwidth]{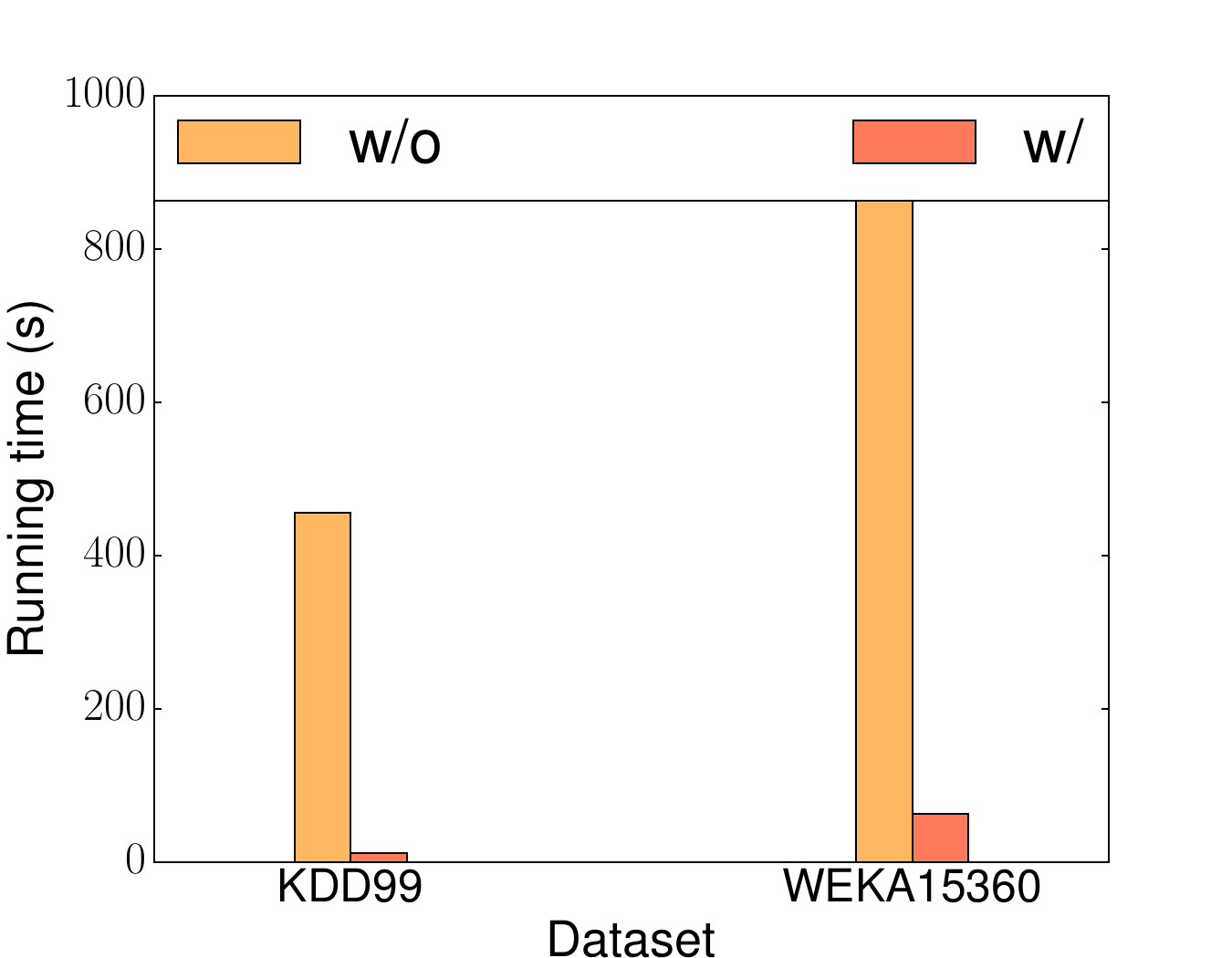}}
\subfigure[SCE]{\includegraphics[width=.24\textwidth]{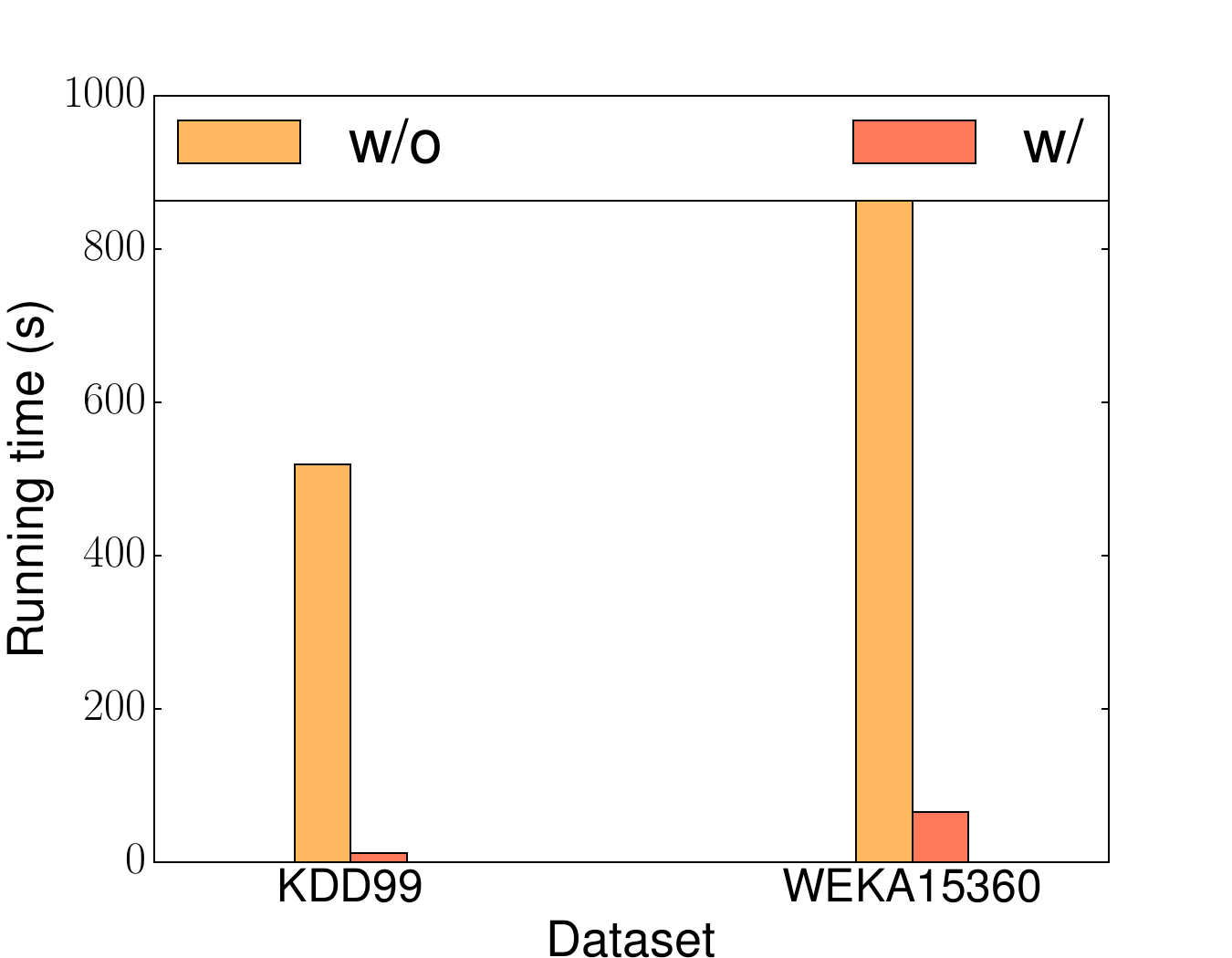}}
\subfigure[LCE]{\includegraphics[width=.24\textwidth]{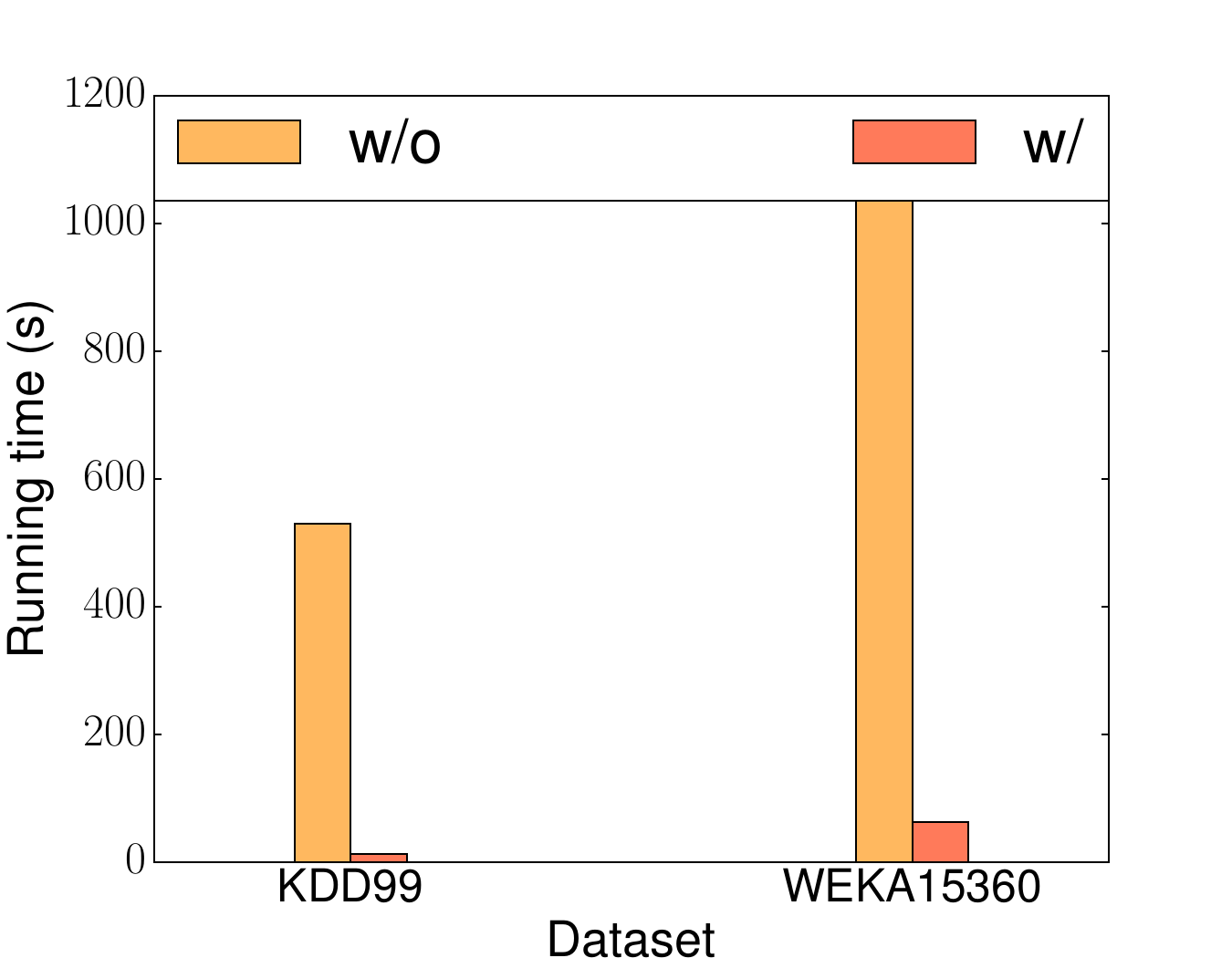}}
\subfigure[CCE]{\includegraphics[width=.24\textwidth]{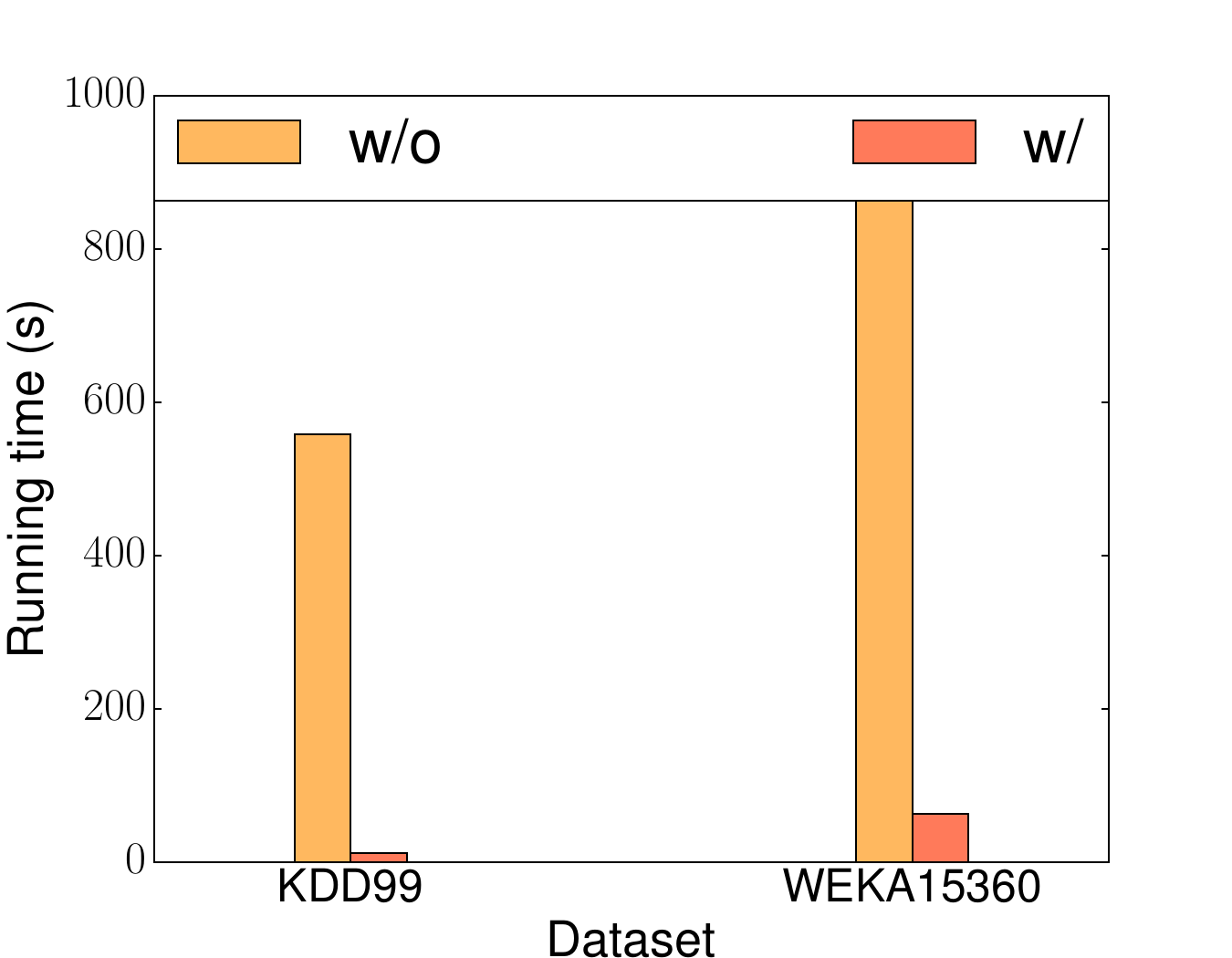}}
\footnotesize
\caption{The running time with or without GrC-based initialization}
\label{fig:EffectGrC}
\end{figure}

In this experiment, we test the effect of GrC-based initialization on Datasets KDD99 and WEKA15360 of Table~\ref{tab:plar:data}, as depicted in Figure~\ref{fig:EffectGrC}. 
We can observe that the running with the GrC-based initialization are extremely less on both datasets KDD99 and WEKA15360 using four different attribute significance measures which demonstrates the GrC-based initialization can efficiently accelerate the whole processing of feature selection. 
\subsubsection{Effect of model parallelism level}\label{sec:high}
\begin{figure}[!htbp]
  \centering
  \includegraphics[width=0.45\textwidth]{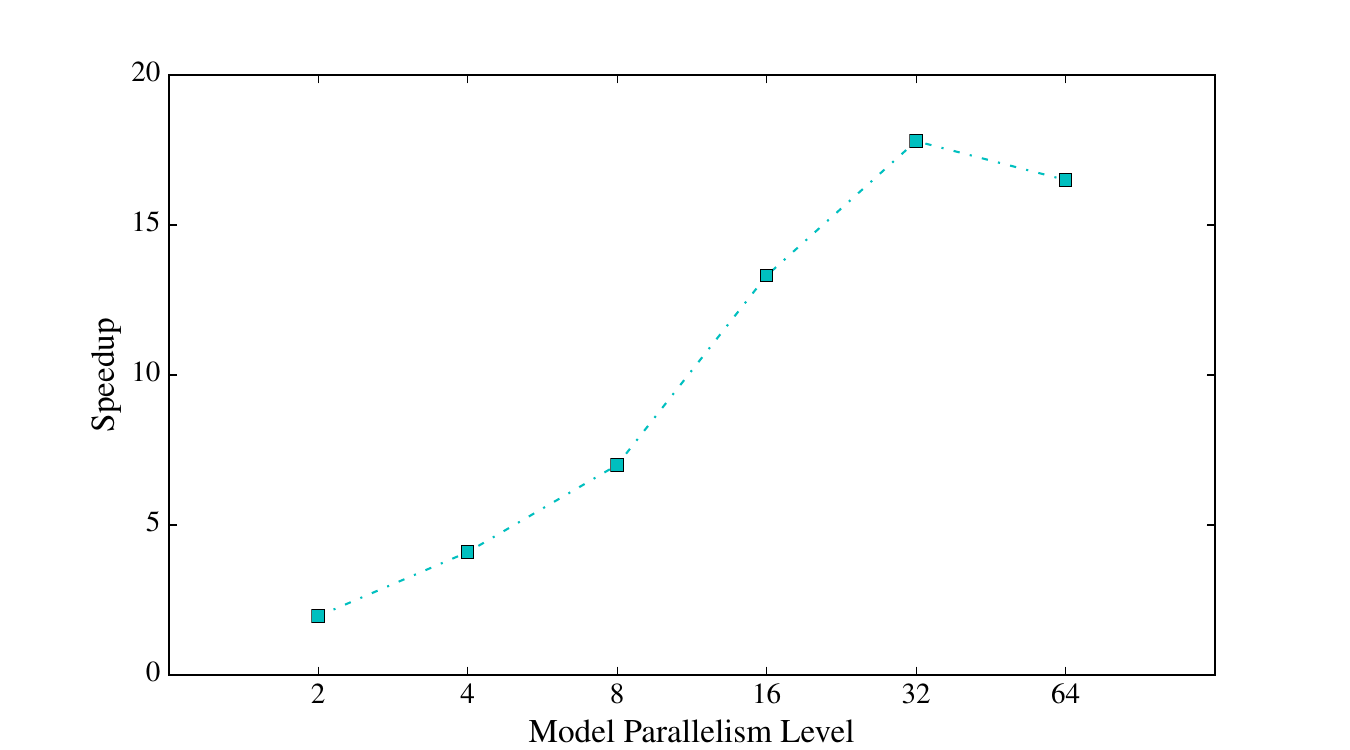}
  \caption{Speedup of varying model parallelism levels}
  \label{fig:gisette_speedup}
\end{figure}

In this section, we test our algorithms with different model parallelism levels for attribute reduction on the dataset Gisette of Table~\ref{tab:plar:data}. We use 64 cores and the attribute measure method SCE for this experiment. 
Gisette is consisted of 5000 features and 6000 samples. 
Suppose that the attribute core is empty, it sequentially evaluates 5000 feature candidates in $1^{\rm{st}}$ iteration, 4999 ones in $2^{\rm{nd}}$ iteration, 4998 ones in $3^{\rm{rd}}$ iteration, $\ldots$, which requires model parallelism more obviously. 
We here set different model parallelism levels: 1, 2, 4, 8, 16, 32 and 64. 
When the model parallelism level is equal to 1, PLAR is degraded into PLAR-DP (only data parallelism). 
We record the first 5 iterations' running time, shown in Table~\ref{tab:Gisette_running_time}. 
When the model parallelism level is 2, PLAR is twice faster than PLAR-DP. As the model parallelism level increases, the running time reduces obviously. 
Intuitively, we employ $\rm{speedup}=\frac{\texttt{running time of PLAR}}{\texttt{running time of PLAR-DP}}$ to show the importance of the model parallelism in PLAR. 
From Figure~\ref{fig:gisette_speedup}, we can clearly observe the speedup is the highest (\textit{i.e.} $17.8\times$) when $\rm{model\ parallelism\ level}= 32$. 

\begin{table}[!htbp]
\tabcolsep 0pt \caption{The running time of each iteration on Gisette (unit: second)}\label{tab:Gisette_running_time} \vspace*{-15pt}
    \begin{flushleft}
    \def\temptablewidth{0.48\textwidth}
        {\rule{\temptablewidth}{1pt}}
        \begin{tabular*}{\temptablewidth}{@{\extracolsep{\fill}}cccccccc}
            \multirow{2}*{Iteration} & \multirow{2}*{PLAR-DP} & \multicolumn{6}{c}{PLAR: model parallelism level} \\
            \cline{3-8}
            & &2&4&8&16&32&64\\
            \hline
1&6262&3080&1570&885&472&350&371\\
2&5975&2982&1480&873&465&343&370\\
3&6261&3059&1497&869&470&344&370\\
4&6115&3017&1484&877&468&344&369\\
5&6194&3155&1512&885&465&348&375\\
\hline
Total &30806&15293&7543&4389&2340&1730&1856
\end{tabular*}
        {\rule{\temptablewidth}{1pt}}
        \end{flushleft}
\end{table}

\section{Conclusions}\label{sec:C}
We have proposed and implemented a highly parallelizable method, PLAR, for large-scale attribute reduction. It consists of GrC-based initialization, model-parallelism and data-parallelism. 
The GrC-based initialization converts the original decision table into a granularity representation which reduce the space complexity from $\mathcal O(|U||A|)$ into $\mathcal O(|U/A||A|)$, and can efficiently accelerate the computation. The model-parallelism is a natural parallel strategy which means we can evaluate all feature candidates at the same time. It becomes much more efficient when there are thousands of features. The data-parallelism means that we can compute a single attribute's significance in parallel which benefits from our decomposition method to the evaluation functions. 
Extensive experimental results show that PLAR is more efficient and scalable than the existing  solutions for large-scale problems. 

\section*{Acknowledgments}
This work is supported by the National Science Foundation of China (Nos. 61573292, 61572406).